\documentclass[preprint,a4paper,10pt,3p]{elsarticle}

\usepackage{amssymb}

\usepackage{lineno}

\newdefinition{rmk}{Remark}

\usepackage{physics}

\usepackage{xcolor}

\usepackage{subfigure}

\journal{Computer Methods in Applied Mechanics and Engineering}

\begin{document}

\begin{frontmatter}


\title{A novel smoothed particle hydrodynamics formulation for thermo-capillary phase change problems with focus on metal additive manufacturing melt pool modeling}

\author[1,2]{Christoph Meier\corref{cor}}
\ead{meier@lnm.mw.tum.de}

\author[1,3]{Sebastian L. Fuchs}
\ead{fuchs@lnm.mw.tum.de}

\author[2]{A. John Hart}
\ead{ajhart@mit.edu}

\author[1]{Wolfgang A. Wall}
\ead{wall@lnm.mw.tum.de}

\cortext[cor]{corresponding author}

\address[1]{Institute for Computational Mechanics, Technical University of Munich, Boltzmannstrasse 15, 85748, Garching, Germany}

\address[2]{Mechanosynthesis Group, Department of Mechanical Engineering, Massachusetts Institute of Technology, 77 Massachusetts Avenue, Cambridge, 02139, MA, USA}

\address[3]{Institute of Continuum and Materials Mechanics, Hamburg University of Technology, Eissendorfer Str. 42, 21073, Hamburg, Germany}

\begin{abstract}
Laser-based metal processing including welding and three dimensional printing, involves localized melting of solid or granular raw material, surface tension-driven melt flow and significant evaporation of melt due to the applied very high energy densities. The present work proposes a weakly compressible smoothed particle hydrodynamics formulation for thermo-capillary phase change problems involving solid, liquid and gaseous phases with special focus on selective laser melting, an emerging metal additive manufacturing technique. Evaporation-induced recoil pressure, temperature-dependent surface tension and wetting forces are considered as mechanical interface fluxes, while a Gaussian laser beam heat source and evaporation-induced heat losses are considered as thermal interface fluxes. A novel interface stabilization scheme is proposed, which is shown to allow for a stable and smooth liquid-gas interface by effectively damping spurious interface flows as typically occurring in continuum surface force approaches. Moreover, discretization strategies for the tangential projection of the temperature gradient, as required for the discrete Marangoni forces, are critically reviewed. The proposed formulation is deemed especially suitable for modeling of the melt pool dynamics in metal additive manufacturing because the full range of relevant interface forces is considered and the explicit resolution of the atmospheric gas phase enables a consistent description of pore formation by gas inclusion. The accuracy and robustness of the individual model and method building blocks is verified by means of several selected examples in the context of the selective laser melting process.\\
\end{abstract}


\end{frontmatter}


\section{Introduction} \label{sec:intro}

The present work proposes a novel smoothed particle hydrodynamics (SPH) formulation for general thermo-capillary phase change problems involving solid, liquid and gaseous phases. A special focus lies on the mesoscale melt pool modeling in metal powder bed fusion additive manufacturing (PBFAM) processes, e.g. selective laser melting (SLM) or electron beam melting (EBM), requiring some additional model constituents that are specific for this application. Since the governing physics are similar, also the melt pool dynamics in laser beam welding (LBW) or electron beam welding (EBW) processes~\cite{Chang2015,Geiger2009,Ki2002a,Ki2002b,Rai2008,Semak1999} lie in the scope of application of the proposed model.

Basically, two main modeling approaches for surface tension effects can be distinguished in the context of SPH: formulations considering the microscale origin of surface tension in form of discrete, phase-dependent inter-particle potentials~\cite{Nugent2000, Tartakovsky2005, Tartakovsky2016} as well as macroscale surface tension models relying on the continuum surface force (CSF) method proposed by Brackbill and Kothe~\cite{Brackbill1996} and widely used also in combination with other spatial discretization schemes such as finite differences, finite volumes or finite elements. The CSF approaches can be further subdivided into formulations that directly discretize the surface tension stress tensor and subsequently determine its divergence as contribution to the discrete momentum equation~\cite{Lafaurie1994, Hu2006} and formulations that rely on the divergence of the continuous surface tension stress tensor resulting in the well-known curvature-proportional surface tension forces in interface normal direction and tangential interface forces proportional to surface tension gradients. The present work will focus on the second category for which the first SPH discretization has been proposed by Morris~\cite{Morris2000}. Subsequently, this formulation has been extended by density-weighted color field gradients~\cite{Adami2010} as well as different interface reconstruction and smoothing techniques~\cite{Andersson2010, Zhang2010, Zhang2015a, Zhang2015b}. There are only very few approaches to incoorporate wetting effects into this type of SPH formulation as e.g. proposed by Breinlinger et al.~\cite{Breinlinger2013} or by Das and Das~\cite{Das2010}. One of the first SPH formulations for thermo-capillary flow, i.e. surface tension effects coupled with a thermal field, has been proposed by Tong and Browne~\cite{Tong2014} and extented by Hopp-Hirschler et al.~\cite{Hopp-Hirschler2018}. {Recently, also several SPH formulations for thermo-capillary phase change problems in the context of PBFAM melt pool modeling have been proposed~\cite{Russell2018,Wessels2018,trautmann2018numerical,Weirather2019,shah2020simulations,furstenau2020generating,dao2021simulations}}. To the best of the authors' knowledge non of the aforementioned thermo-capillary SPH formulations has incorporated wetting effects so far, which are expected, however, to play an important role on the length scales relevant for metal PBFAM.

In metal PBFAM, a focused laser beam, typically within an inert gas atmosphere, melts pre-defined contours into thin layers of pre-applied metal powder to create the cross-section of a final solid part in a repeated layer-wise buildup procedure. Under typical processing conditions the peak temperatures on the melt pool surface exceed the boiling temperature of the liquid metal. The density jump and accompanied recoil pressure in the phase transition from liquid metal to metal vapor results in a considerable distortion and highly dynamic topology changes of the liquid-gas interface at the melt pool surface giving rise to defects such as spatter, i.e. ejection of melt drops, or pores, i.e. gas bubble inclusions~\cite{Meier2017}. Pioneering modeling approaches in this field are e.g. given by the thermo-hydrodynamics finite element model proposed by Khairallah et al.~\cite{Khairallah2014, Khairallah2016, Khairallah2020}, who considered temperature-dependent surface tension and evaporation-induced recoil pressure forces, based on a phenomenological recoil pressure model~\cite{Anisimov1995}, as primary driving forces of the process. Comparable models based on finite difference, finite volume, finite element, Lattice Boltzmann or meshfree discretizations are e.g. given by~\cite{Geiger2009,Russell2018, Weirather2019, Lee2015, Leitz2018, Qiu2015, Panwisawas2017, Wessels2018, Otto2012, Yan2018, egorov2020, Gurtler2013,Yu2016,Yuan2015}. A more refined model has been proposed by~\cite{Tan2013, Tan2014, Kouraytem2019}, where the gas / vapor phase is explicitly resolved. Typically, the aforementioned models do not account for wetting effects at the triple line solid-liquid-gas. On the contrary, the works~\cite{Korner2011, Korner2013, Markl2015} specifically focus on the interplay between wetting effects and different power particle configurations, without considering however evaporation-induced recoil pressure.

The present work proposes a weakly compressible SPH formulation for thermo-capillary phase change problems involving solid, liquid and gaseous phases. Specifically, evaporation-induced recoil pressure, temperature-dependent surface tension and wetting forces are considered as liquid-gas interface fluxes in the Navier Stokes equation. In the thermal problem, a Gaussian laser beam heat source as well as evaporation-induced heat losses are considered as liquid-gas interface fluxes, while convection boundary conditions are obsolete due to the explicit modeling of the atmospheric gas phase. All mechanical and thermal interface fluxes are modeled in a diffuse sense in analogy to the CSF approach. The following original contributions of the present work can be identified: The first SPH formulation for thermo-capillary problems is proposed that also considers wetting effects. A novel interface stabilization scheme based on viscous interface forces is proposed, which is shown to allow for a stable and smooth liquid-gas interface by effectively damping spurious interface flows well-known for the CSF approach. Moreover, different SPH discretizations for the tangential projection of the temperature gradient, as required for the discrete Marangoni forces, are reviewed. Based on a thorough analysis it is shown that standard \textit{two-sided} gradient approximations are sufficient for this purpose as long as zero-order consistency is satisfied, e.g. by anti-symmetric gradient construction. In the context of metal AM melt pool modeling, the present approach is - to the best of the authors' knowledge - the first model that i) considers the full range of relevant interface forces consisting of evaporation-induced recoil pressure, temperature-dependent surface tension and wetting forces, and ii) resolves the atmospheric gas phase and, thus, can consistently account for defects such as gas inclusions.

The remainder of this work is organized as follows: Section~\ref{sec:goveq} presents the governing equations, i.e. continuity equation, momentum equation, energy equation and equation of state, in space-continuous form. Discretization in space, based on SPH, and in time, based on an explicit velocity-Verlet scheme, is presented in Sections~\ref{sec:nummeth_sph} and~\ref{sec:nummeth_sph_timint}. In Section~\ref{sec:numex_tantempgrad} different SPH approximations for the tangential temperature gradient are thoroughly analyzed and compared. {Finally, in Section~\ref{sec:numex}, the accuracy of the individual model and method components is verified by means of selected benchmark examples with analytical/numerical reference solutions. Eventually, the suitability of the proposed melt pool model for typical metal AM application scenarios is verified  by means of point and line melting examples with and without resolved powder particles. Here, a special focus lies on the robustness of the computational model, i.e. the ability to represent challenging and practically relevant scenarios of dynamically changing interface topologies (e.g. generation of melt spatter or gas inclusions) without inducing spurious interface flows or instabilities of the discretization scheme.}

\section{Governing Equations} \label{sec:goveq}

Throughout this work two-phase flow problems of a liquid phase $\Omega^{l}$ and a gas phase $\Omega^{g}$ are considered that interact with a solid phase $\Omega^{s}$ and allow for reversible phase transition between liquid and solid phase. The overall problem domain splits according to  $\Omega = \Omega^{l} \cup \Omega^{g} \cup \Omega^{s}$ and the two-phase fluid domain is given by $\Omega^{f} = \Omega^{l} \cup \Omega^{g}$. In the context of metal AM melt pool modeling the solid, liquid and gas phase correspond to the solid metal, the molten metal and the atmospheric gas in the build chamber 
of an AM device.

\subsection{Fluid phases} \label{subsec:goveq_fluid}
The liquid and gas phase are governed by the  \textit{weakly compressible}, instationary and anisothermal Navier-Stokes equations in the domain~$\Omega^{f}=\Omega^{l} \cup \Omega^{g}$. The problem shall be described by the continuity equation
\begin{equation} \label{eq:fluid_conti}
\dv{\rho}{t} = -\rho \div \vectorbold{u} \qin \Omega^{f},
\end{equation}
the Navier-Stokes momentum equation
\begin{equation} \label{eq:fluid_momentum}
\dv{\vectorbold{u}}{t} = \frac{1}{\rho} \left(-\grad{p} + \vectorbold{f}_{\nu} + \tilde{\vectorbold{f}}^{{lg}}_{s} + \tilde{\vectorbold{f}}^{{slg}}_{w} + \tilde{\vectorbold{f}}^{{lg}}_{v} \right) + \vectorbold{g} \qin \Omega^{f},
\end{equation}
as well as the energy equation:
\begin{equation} \label{eq:fluid_energy}
c _p\dv{T}{t} =  \frac{1}{\rho} \left(-\div \vectorbold{q} + \tilde{s}^{lg}_v + \tilde{s}^{lg}_l \right)  \qin \Omega.
\end{equation}
Following a weakly compressible approach, density~$\rho$ and pressure~$p$ are linked via the equation of state
\begin{equation} \label{eq:fluid_eos}
p\qty(\rho) = c^{2} \qty(\rho - \rho_{0}) = p_{0} \qty(\frac{\rho}{\rho_{0}} - 1)  \qin \Omega^{f},
\end{equation}
which closes the system of equations for the six unknowns velocity $\vectorbold{u}$ (three components), density~$\rho$, pressure~$p$ and temperature~$T$. The individual contributions to these equations 
will be discussed in the following.

\subsubsection{Momentum equation} \label{subsec:goveq_fluid_momentum}
In equation~\eqref{eq:fluid_momentum}, contributions from viscous forces~$\vectorbold{f}_{\nu}$, surface tension forces~$\tilde{\vectorbold{f}}^{{lg}}_{s}$, wetting forces~$\tilde{\vectorbold{f}}^{{slg}}_{w}$ and evaporation-induced recoil pressure forces $\tilde{\vectorbold{f}}^{{lg}}_{v}$, each per unit volume, as well as body forces~$\vectorbold{g}$ per unit mass, can be identified. For incompressible Newtonian fluids the viscous forces read
\begin{equation}
\vectorbold{f}_{\nu} = \eta \laplacian{\vectorbold{u}},
\end{equation}
with dynamic viscosity~$\eta$. Following the continuum surface force (CSF) approach by Brackbill and Kothe~\cite{Brackbill1996} we consider surface tension and wetting effects in the momentum equation~\eqref{eq:fluid_momentum} as volumetric forces distributed across an interfacial volume of finite width instead of additional boundary conditions at the liquid-gas interface area and the triple line 
solid-liquid-gas. In the following, these interface forces are marked by a tilde symbol and by a superscript indicating the relevant interface, e.g. $\tilde{\vectorbold{f}}^{{lg}}$ for forces on the 2D liquid-gas interface or $\tilde{\vectorbold{f}}^{{slg}}$ for forces on the 1D solid-liquid-gas interface (triple line). Specifically, the distributed surface tension forces consist of the following two contributions in interface normal and tangential direction
\begin{equation} \label{eq:fluid_surfacetension}
\tilde{\vectorbold{f}}^{{lg}}_{s} = -\alpha \kappa \vectorbold{n}^{lg} \delta^{lg} + \left(\vectorbold{I} -  \vectorbold{n}^{lg} \otimes \vectorbold{n}^{lg} \right) \grad{\alpha} \delta^{lg},
\end{equation}
with the surface tension coefficient $\alpha$, the interface curvature $\kappa:=\div \vectorbold{n}^{lg}$, the liquid-gas interface normal $\vectorbold{n}^{lg}:= \grad{c^{lg}}/||\grad{c^{lg}}||$, the phase-specific color field $c^{lg}$ between the liquid and gas phase, to be defined in Section~\ref{subsec:nummeth_sph_colorfield}, as well as the surface delta function $\delta^{lg}:=||\grad{c^{lg}}||$ between liquid and gas phase. The surface delta function is employed to distribute interface surface forces across interface domains of finite thickness. It is non-zero only on these interface domains and its integral over the interface thickness direction is normalized to one (see also Section~\ref{subsec:nummeth_sph_colorfield}). Throughout this work a purely temperature-dependent surface tension coefficient, i.e. $\grad{\alpha}= \alpha'(T) \grad{T}$ with $\alpha'(T)=d \alpha(T) / d T$, is considered. Specifically, a linear temperature-dependence of the surface tension is considered in the examples in Section~\ref{sec:numex} according to
\begin{equation} \label{eq:fluid_surfacetension_tempdepend}
\alpha(T) = \alpha_0 -  \alpha'_0 (T-T_{\alpha_0}),
\end{equation}
where $\alpha_0$ is the surface tension at reference temperature $T_{\alpha_0}$. Moreover, the wetting forces~{\cite{Breinlinger2013}} are given by
\begin{equation} \label{eq:fluid_wetting}
\tilde{\vectorbold{f}}^{{slg}}_{w} = \alpha \left( \cos \theta - \cos \theta_0 \right)  \vectorbold{t}^{sf} \delta^{lg} \delta^{sf} ,
\end{equation}
with the equilibrium wetting angle $\theta_0$ and the current wetting angle $\theta$ defined via $\cos \theta := \vectorbold{n}^{lg} \cdot \vectorbold{n}^{sf}$. Here, the solid-fluid interface normal vector between the domains $\Omega^{s}$ and $\Omega^{f}$ is defined as $\vectorbold{n}^{sf}:= \grad{c^{sf}}/||\grad{c^{sf}}||$ on the basis of a phase-specific color field $c^{sf}$ between the solid and fluid phase to be defined in Section~\ref{subsec:nummeth_sph_colorfield}. Similar to the liquid-gas interface, also the surface delta function of the solid-fluid interface follows the relation $\delta^{sf}:=||\grad{c^{sf}}||$. Moreover, the tangent vector $\vectorbold{t}^{sf}$ is defined as the projection of the liquid-gas interface normal vector $\vectorbold{n}^{lg}$ onto the solid-fluid interface surface, {defined by its normal vector $\vectorbold{n}^{sf}$~\cite{Breinlinger2013}:}
\begin{equation} \label{eq:fluid_wetting2}
\vectorbold{t}^{sf} = \frac{\vectorbold{n}^{lg} - (\vectorbold{n}^{lg} \cdot \vectorbold{n}^{sf}) \vectorbold{n}^{sf}}{|| \vectorbold{n}^{lg} - (\vectorbold{n}^{lg} \cdot \vectorbold{n}^{sf}) \vectorbold{n}^{sf} ||}.
\end{equation}
Besides these standard capillary force contributions, the high peak temperatures at the melt pool surface at typical metal AM process conditions give rise to considerable evaporation effects. As common in the modeling of these processes, a phenomenological model for the evaporation-induced recoil pressure forces acting on the melt pool surface according to the work by Anisimov~\cite{Anisimov1995} is employed:
\begin{equation} \label{eq:fluid_recoil}
\tilde{\vectorbold{f}}^{{lg}}_{v} = -p_v(T) \vectorbold{n}^{lg} \delta^{lg} \quad \text{with} \quad p_v(T) = C_P \exp \left[ - C_T \left( \frac{1}{T} - \frac{1}{T_v} \right)\right],
\end{equation}
where the constants $C_P = 0.54 p_a$ and $C_T=\bar{h}_v/R$ contain the atmospheric pressure $p_a$, the molar latent heat of evaporation $\bar{h}_v$ and the molar gas constant $R$. Moreover, $T_v$ is the boiling temperature.

\begin{rmk}
The working principle of the employed phenomenological evaporation model relies on local peak temperatures in the melt pool that exceed the boiling temperature of the liquid metal. In scenarios with very high laser powers the resulting peak temperatures might lead to negative surface tension coefficients if~\eqref{eq:fluid_surfacetension_tempdepend} is applied without further correction. Thereto, in this work a regularized version of~\eqref{eq:fluid_surfacetension_tempdepend} is applied which keeps the surface tension coefficient $\alpha(T)$ constant if it has already decreased to $10 \%$ of its reference value $\alpha_0$ and the temperature is further increasing.
\end{rmk}

\subsubsection{Energy equation} \label{subsec:goveq_fluid_energy}

The energy equation~\eqref{eq:fluid_energy} contains the mass-specific heat capacity $c_p$, the heat flux $\vectorbold{q}:= -k \grad{T}$ according to Fourier's law with thermal conductivity $k$ as well as heat fluxes stemming from the laser beam heat source $\tilde{s}^{lg}_l$ and from evaporation-induced heat losses $\tilde{s}^{lg}_v$, each per unit volume. The former is given by
\begin{equation} \label{eq:fluid_heatsource}
\tilde{s}^{lg}_l = \chi_l <\! -\vectorbold{n}^{lg} \! \cdot \! \vectorbold{e}_{l} \! > \, s^{lg}_l(\vectorbold{x}) \, \delta^{lg} \quad \text{with} \quad s^{lg}_l(\vectorbold{x})=s^{lg}_{l0} \, \exp \! \left[- 2 \left( \frac{||\vectorbold{x}-\vectorbold{x}_0||}{r_w} \right)^2 \right],
\end{equation}
where the Macauley bracket $<...>$ returns the value of its argument if the argument is positive and zero otherwise. The irradiance $s^{lg}_l(\vectorbold{x})$ describes the incident laser power per unit area at position $\vectorbold{x}$ as a function of the laser beam center position $\vectorbold{x}_0$ and has the form of a Gauss distribution, from which $s^{lg}_{l0}$ is the peak value and $r_{w}=2 \sigma$ represents two times the standard deviation $\sigma$. The corresponding diameter $d_w=2 r_{w}$ is a frequently used measure for the effective laser beam diameter. In addition,  $\vectorbold{e}_{l}$ is the unit vector representing the laser beam direction and $\chi_l$ the laser energy absorptivity. Eventually, following the same phenomenological model as for the recoil pressure~\eqref{eq:fluid_recoil}, the evaporation-induced heat loss reads
\begin{equation} \label{eq:fluid_evaporation}
\tilde{s}^{lg}_v = s^{lg}_v \, \delta^{lg} \quad \text{with} \quad s^{lg}_v = - \dot{m}^{lg}_v [h_v+ h(T)], \quad \dot{m}^{lg}_v = 0.82 \, c_s \, p_v(T) \, \sqrt{\frac{C_M}{T}}, \quad h(T)=\int \limits_{T_{h,0}}^T  c_p \, \, d\bar{T},
\end{equation}
where the enthalpy rate per unit area $s^{lg}_v$ results from the vapor mass flow per unit area $\dot{{m}}^{lg}_v$ at the melt pool surface and the sum of the specific enthalpy $h(T)$ and the latent heat of evaporation $h_v$, both per unit mass. Moreover, $T_{h,0}$ is a reference temperature of the specific enthalpy and the constant $C_M=M/(2\pi R)$ contains the molar mass $M$ and the molar gas constant $R$. Finally, $p_v(T)$ is the recoil pressure defined in~\eqref{eq:fluid_recoil} and $c_s$ the so-called sticking constant which takes on a value close to one, i.e. $c_s \approx 1$ for metals~\cite{Khairallah2016, Weirather2019}.

\begin{rmk}
According to equation~\eqref{eq:fluid_heatsource} the laser beam heat source is applied at the liquid-gas interface, which is characterized by $\delta^{lg} \neq 0$. In the proposed model, the laser heat source also acts on solid surfaces, i.e. at solid-gas interfaces characterized by $\delta^{sg} \neq 0$. Here, the surface delta function $\delta^{sg}$ of the solid-gas interface is defined in analogy to the surface delta function $\delta^{lg}$ of the liquid-gas interface.
\end{rmk}

\subsubsection{Equation of state} \label{subsec:goveq_fluid_eos}

In the equation of state~\eqref{eq:fluid_eos} the reference density~$\rho_{0}$, which equals the initial density throughout this work, the reference pressure~$p_{0} = \rho_{0} c^{2}$ and the artificial speed of sound~$c$ can be indentified. Note that the commonly applied weakly compressible approach only represents deviations from the reference pressure, i.e., $p\qty(\rho_{0}) = 0$, and not the total pressure. In order to limit density fluctuations to an acceptable level, while still avoiding too severe time step restrictions, Morris et al.~\cite{Morris1997} discussed strategies on how to determine an appropriate value of the artificial speed of sound.

\subsection{Solid phase} \label{subsec:goveq_solid}

Since the focus of this work lies on melt pool thermo-hydrodynamics, the assumption of a rigid and immobile solid phase (substrate and powder grains in the context of PBFAM processes), which is typical for 
mesoscale PBFAM models, is made. Thus, only the energy equation~\eqref{eq:fluid_energy} is solved for the solid phase.

\subsection{Phase transition} \label{subsec:goveq_phase_transition}

As presented in Section~\ref{sec:nummeth_sph}, the spatial discretization will be based on smoothed particle hydrodynamics (SPH). Due the Lagrangian nature of this scheme, each (material) particle directly carries its phase information. Based on this information, the corresponding field equations with phase-specific parameter values are evaluated for each particle. Material particles undergo the phase transition \textit{solid} $\leftrightarrow$ \textit{liquid} when passing the melt temperature $T_m$. Since the vapor phase is not modeled explicitly, the phase transition \textit{liquid} $\leftrightarrow$ \textit{vapor} is only considered implicitly in terms of evaporation-induced recoil pressure forces~\eqref{eq:fluid_recoil} and heat losses~\eqref{eq:fluid_evaporation}. While the latent heat of evaporation is already contained~\eqref{eq:fluid_evaporation}, the latent heat of melting could be considered in a straightforward manner as well by employing e.g. an apparent capacity scheme relying on an increased heat capacity $c_p$ within a finite temperature interval~\cite{proell2020phase}. However, for simplicity temperature-independent parameter values are considered in the present work (except for the surface tension coefficient).

\subsection{Initial and boundary conditions}

In general, the system of partial differential equations~\eqref{eq:fluid_conti}-\eqref{eq:fluid_energy} is subject to the following initial conditions
\begin{equation}
\rho = \rho_0, \quad \vectorbold{u} = \vectorbold{u}_{0}, \quad T=T_0 \qin \Omega \qq{at} t = 0.
\end{equation}
Throughout this work, only systems that are initially in static equilibrium, i.e. $\vectorbold{u}_{0}=\vectorbold{0}$, are considered.
In addition, Dirichlet and Neumann boundary conditions are required on the domain boundary $\Gamma = \partial\Omega$:
\begin{equation}
\vectorbold{u} = \hat{\vectorbold{u}} \qq{on} \Gamma^{\vectorbold{u}}_{D}, \quad \quad
\vectorbold{t} = \hat{\vectorbold{t}} \qq{on} \Gamma^{\vectorbold{u}}_{N}, \quad \quad
T = \hat{T} \qq{on} \Gamma^T_{D}, \quad \quad
\vectorbold{q} = \hat{\vectorbold{q}} \qq{on} \Gamma^T_{N} \, ,
\end{equation}
with boundary velocity~$\hat{\vectorbold{u}}$, boundary traction~$\hat{\vectorbold{t}}$, boundary temperature $T$ and boundary heat flux $\hat{\vectorbold{q}}$ on the Dirichlet and Neumann boundaries $\Gamma = \Gamma^{\vectorbold{u}}_{D} \cup \Gamma^{\vectorbold{u}}_{N}$ and $\Gamma^{\vectorbold{u}}_{D} \cap \Gamma^{\vectorbold{u}}_{N} = \emptyset$ as well as $\Gamma = \Gamma^T_{D} \cup \Gamma^T_{N}$ and $\Gamma^T_{D} \cap \Gamma^T_{N} = \emptyset$.

\section{Spatial discretization via smoothed particle hydrodynamics} \label{sec:nummeth_sph}
\subsection{Approximation of field quantities via smoothing kernel} \label{subsec:nummeth_sph_kernel}
The fundamental concept of SPH is based on the approximation of a field quantity~$f$ via a smoothing operation and on the discretization of the domain~$\Omega$ with discretization points following the fluid motion and therefore being denoted as particles. Introducing a smoothing kernel~$W\qty(r,h)$ (see e.g.~\cite{Liu2010,Monaghan2005}) leads to the following sequence of approximations of an arbitrary field quantity~$f$: 
\begin{equation}
\label{eq:sph_approximation}
f\qty(\vectorbold{r}) \approx \int_{\Omega} f\qty(\vectorbold{r}') W\qty(\qty| \vectorbold{r} - \vectorbold{r}' |, h) \dd{\vectorbold{r}'} \approx \sum_{j} V_{j} f\qty(\vectorbold{r}_{j}) W\qty(\qty| \vectorbold{r} - \vectorbold{r}_{j} |, h),
\end{equation}
commiting a \textit{smoothing error} in the first and an \textit{integration error} in the second approximation step~\cite{Quinlan2006}. The smoothing kernel~$W\qty(r,h)$ is a monotonically decreasing, smooth function that dependents on the distance~$r$ from the kernel center and a smoothing length~$h$. The smoothing length~$h$ together with a scaling factor~$\kappa$ define the support radius of the smoothing kernel~$r_{c} = \kappa h$. The Dirac delta function property $\lim_{h \rightarrow 0}{ W\qty(r, h) } = \delta\qty(r)$ ensures an exact representation of a field quantity~$f$ in the limit $h \rightarrow 0$. Compact support, i.e., $W\qty(r, h) = 0$ for $r > r_{c}$, as well as positivity, i.e., $W\qty(r, h) \geq 0$ for $r \leq r_{c}$, are typical properties of standard smoothing kernels~$W\qty(r,h)$. In addition, the normalization property requires that $\int_{\Omega} W\qty(\qty| \vectorbold{r} - \vectorbold{r}' |, h) \dd{\vectorbold{r}'} = 1$. In the second approximation step, the domain integral is replaced by a summation over discrete volumes~$V_{j}$ located at the positions of the material discretization points (particles)~$j$. A straightforward approximation for  the gradient of quantity~$f$ follows directly by differentiation of \eqref{eq:sph_approximation}:
\begin{equation}
\label{sph_general_gradient}
\grad{f}\qty(\vectorbold{r}) \approx \int_{\Omega} f\qty(\vectorbold{r}') \grad{W \qty(\qty| \vectorbold{r} - \vectorbold{r}' |, h)} \dd{\vectorbold{r}'} \approx \sum_{j} V_{j} f\qty(\vectorbold{r}_{j}) \grad{W \qty(\qty| \vectorbold{r} - \vectorbold{r}_{j} |, h)},
\end{equation}
Note that this (simple) gradient approximation shows some particular disadvantages, hence, more advanced approximations for gradients are given in the literature \cite{Monaghan2005} and will also be applied in the following. Applying these gradient approximations reduces the partial differential equations~\eqref{eq:fluid_conti} and~\eqref{eq:fluid_energy} to ordinary differential equations that are solved, i.e., evaluated and integrated in time, for all particles~$i$ in the domain~$\Omega$ (cf. Sections \ref{subsec:nummeth_sph_momentum}-\ref{sec:nummeth_sph_timint}). As a result, all fluid quantities are evaluated at and associated with particle positions, meaning each particle carries its corresponding fluid quantities.

\begin{rmk}
In the following, a quantity~$f$ evaluated for particle~$i$ at position~$\vectorbold{r}_{i}$ is written as~$f_{i} = f\qty(\vectorbold{r}_{i})$. In addition, the short notation $W_{ij} = W\qty(r_{ij}, h)$ denotes the smoothing kernel~$W$ evaluated for particle~$i$ at position~$\vectorbold{r}_{i}$ with neighboring particle~$j$ at position~$\vectorbold{r}_{j}$, where $r_{ij} = \qty|\vectorbold{r}_{ij}| = \qty|\vectorbold{r}_{i} - \vectorbold{r}_{j}|$ is the absolute distance between particles~$i$ and~$j$. Similarly, the derivative of the smoothing kernel~$W$ with respect to the absolute distance~$r_{ij}$ is denoted by $\pdv*{W}{r_{ij}} = \pdv*{W\qty(r_{ij}, h)}{r_{ij}}$.
\end{rmk}

\begin{rmk}
In this work, a quintic spline smoothing kernel~$W\qty(r, h)$ is considered as defined in \cite{Morris1997} with smoothing length~$h$ and compact support with support radius~$r_{c} = \kappa h$ and scaling factor~$\kappa = 3$.
\end{rmk}

\subsection{Initial particle spacing} \label{subsec:nummeth_sph_spacing}
Within this contribution, the domain $\Omega$ is initially filled with particles located on a regular grid with particle spacing~$\Delta{}x$, thus in $d$ dimensional space each particle initially occupies an effective volume of~$\qty(\Delta{}x)^{d}$. The mass of a particle~$i$ is then set using the reference density according to $m_{i} = \rho_{0} \qty(\Delta{}x)^{d}$ and remains constant throughout the simulation. In general, the initial particle spacing~$\Delta{}x$ can be freely chosen, however, within this work the initial particle spacing~$\Delta{}x$ is set equal to the smoothing length~$h = \flatfrac{r_{c}}{\kappa}$.
\subsection{Density summation} \label{subsec:nummeth_sph_summation}
The density of a particle~$i$ is determined via summation of the respective smoothing kernel contributions of all neighboring particles~$j$ within the support radius~$r_{c}$
\begin{equation} \label{eq:sph_densum}
\rho_{i} = m_{i} \sum_{j} W_{ij} \, .
\end{equation}
This approach is typically denoted as density summation and results in an exact conservation of mass in the fluid domain, which can be shown in a straight-forward manner considering the commonly applied normalization of the smoothing kernel to unity. It shall be noted that the density field may alternatively be obtained via SPH discretization and time integration of the continuity equation~\eqref{eq:fluid_conti}, cf. Liu and Liu~\cite{Liu2010}.
\subsection{Momentum equation} \label{subsec:nummeth_sph_momentum}
Following the standard SPH discretization procedure the discrete version of~\eqref{eq:fluid_momentum} can be formulated as
\begin{equation} \label{eq:sph_momentum}
\vectorbold{a}_{i} = \frac{1}{m_{i}} \left[ \vectorbold{F}_{p,i} + \vectorbold{F}_{\nu,i} + \vectorbold{F}_{s,i} + \vectorbold{F}_{w,i} + \vectorbold{F}_{v,i} + \vectorbold{F}_{d,i} \right] + \vectorbold{b}_{i}
\, ,
\end{equation}
where $\vectorbold{a}_{i} = \dv*{\vectorbold{u}_{i}}{t}$ represents the total acceleration of particle~$i$ whereas the pressure force $\vectorbold{F}_{p,i}$, viscous force $\vectorbold{F}_{\nu,i}$, surface tension force $\vectorbold{F}_{s,i}$, wetting force $\vectorbold{F}_{w,i}$ as well as vapor-induced recoil pressure force $\vectorbold{F}_{v,i}$ acting on particle~$i$ result from summation of all particle-to-particle interaction contributions with neighboring particles~$j$. Optionally, additional viscous dissipation forces $\vectorbold{F}_{d,i}$ are applied at the interfaces solid-liquid and liquid-gas, which will motivated and further detailled in Section~\ref{subsec:nummeth_sph_dissipation}. The pressure and viscous forces in the momentum equation~\eqref{eq:sph_momentum} are discretized following a formulation proposed by Adami et al.~\cite{Adami2012, Adami2013}:
\begin{equation} \label{eq:sph_momentum_pressure_and_viscous}
\vectorbold{F}_{p,i} + \vectorbold{F}_{\nu,i} = \sum_{j} \qty(V_{i}^{2}+V_{j}^{2}) \qty[ - \bar{p}_{ij} \pdv{W}{r_{ij}} \vectorbold{e}_{ij} + \bar{\eta}_{ij} \frac{\vectorbold{u}_{ij}}{r_{ij}} \pdv{W}{r_{ij}} ]\, ,
\end{equation}
with volume~$V_{i} = m_{i}/\rho_{i}$ of particle~$i$, unit vector $\vectorbold{e}_{ij} = \flatfrac{\vectorbold{r}_{i} - \vectorbold{r}_{j}}{\qty|\vectorbold{r}_{i} - \vectorbold{r}_{j}|} = \flatfrac{\vectorbold{r}_{ij}}{r_{ij}}$, relative velocity $\vectorbold{u}_{ij} = \vectorbold{u}_{i}-\vectorbold{u}_{j}$ as well as inter-particle averaged pressure and dynamic viscosity:
\begin{equation} \label{eq:sph_mom_wght_press_and_visc}
\bar{p}_{ij} = \frac{\rho_{j}p_{i}+\rho_{i}p_{j}}{\rho_{i} + \rho_{j}} \, , \quad \quad \bar{\eta}_{ij} = \frac{2\eta_{i}\eta_{j}}{\eta_{i}+\eta_{j}} \, .
\end{equation}
Also the transport velocity formulation proposed in~\cite{Adami2013}, which utilizes a constant background pressure $p_b$ to suppress the problem of tensile instability, is employed in the present work. For the sake of briefity, the definition of the modified advection velocity and the additional terms in the momentum equation stemming from the aforementioned transport velocity contribution are not further delineated and the reader is kindly referred to the original publication~\cite{Adami2013}. The accuracy and stability of this formulation has readily been demonstrated on the basis of well-known benchmark tests in the context of computational fluid dynamics~\cite{Adami2013} and fluid-structure interaction~\cite{Fuchs2020}. The remaining force contributions will be discussed in the following.
\subsubsection{Discretization of phase interfaces} \label{subsec:nummeth_sph_colorfield}
The interface forces to be defined in the following rely on a representation of the different phase domains and interfaces via a color field function. Here, we define the color field  according to
\begin{equation} \label{eq:sph_colorfield}
c_i^j = \left\{\begin{array}{ll}
                                 1 , & \text{if particles i and j belong to different phases}\\
                                0, & \text{else}
                    \end{array}\right.
\end{equation}
 as well as the density-weighted color field gradient according to
 \begin{equation} \label{eq:sph_colorfieldgradient}
\grad{c}_i = \frac{1}{V_{i}} \sum_{j} \qty(V_{i}^{2}+V_{j}^{2}) \bar{c}_{ij} \pdv{W}{r_{ij}} \vectorbold{e}_{ij} 
\quad \text{with} \quad \bar{c}_{ij} = \frac{\rho_j}{\rho_i+\rho_j} c_i^i + \frac{\rho_i}{\rho_i+\rho_j} c_j^i
\end{equation}
following the approach proposed by Adami et al.~\cite{Adami2010}. Note that $ c_i^i \equiv 0$ according to~\eqref{eq:sph_colorfield}. Based on the definition of the color field gradient 
the interface normal and the surface delta function of particle $i$ read
\begin{equation} \label{eq:sph_normalanddelta}
\vectorbold{n}_{i} = \frac{\grad{c}_i}{|| \grad{c}_i ||} \quad \text{and} \quad \delta_i = || \grad{c}_i ||.
\end{equation}
Note that this procedure leads to an \textit{outward-pointing} interface normal vector with respect to the phase of particle $i$. Moreover, it has to be emphasized that these metrics are exclusively used to define the interface between two phases. Consequently, in the calculations according to~\eqref{eq:sph_colorfield}-\eqref{eq:sph_normalanddelta} only two different phases are distinguished (and \textit{not} three independent phases which might occur at the triple line solid-liquid-gas). Specifically, the liquid-gas interface (superscript $"lg"$) is defined by only considering particles of the liquid and the gas phase (i.e. no contribution of solid particles). The solid-gas interface (superscript $"sg"$) is defined by only considering particles of the solid and the gas phase. The solid-fluid interface (superscript $"sf"$) is defined by considering all particles but only distinguishing between either particles of the solid phase or the (combined) fluid phase (sum of particles from the liquid and gas phase).

\begin{rmk}
In case of high density ratios between two phases the definition of $\delta_i$ according to~\eqref{eq:sph_colorfieldgradient}-\eqref{eq:sph_normalanddelta} ensures that the majority of a flux contribution (i.e. of mechanical interface forces or thermal heat fluxes) distributed over the interface via $\delta_i$ acts on the side of the interface associated with the \textit{heavier} phase. Thus, in the melt pool examples in Section~\ref{sec:numex_am}, where typical density ratios $\rho_l / \rho_g > 100$ between melt and gas phase are considered, more than $99\%$ of a flux term is carried by the melt phase. Therefore, for simplicity only the contributions to the melt phase are considered for the mechanical and thermal flux terms in~\eqref{eq:fluid_surfacetension},~\eqref{eq:fluid_wetting},~\eqref{eq:fluid_recoil},~\eqref{eq:fluid_heatsource} and~\eqref{eq:fluid_evaporation}. This approximation seems reasonable when considering typical parameter uncertainties and accuracy requirements in AM melt pool modeling. However, it is emphasized that this procedure is by no means an inherent limitation of the proposed methodology and can easily be changed.
\end{rmk}

\subsubsection{Surface tension forces} \label{subsec:nummeth_sph_surface tension}

In the following, the surface tension forces are split into two contributions $\vectorbold{F}_{s,i}=\vectorbold{F}_{s \kappa,i}+\vectorbold{F}_{s m,i}$, with $\vectorbold{F}_{s \kappa,i}$ representing the curvature-proportional surface tension normal forces and $\vectorbold{F}_{s m,i}$ representing the tangential Marangoni forces due to surface tension gradients. The first contribution is given by
\begin{equation} \label{eq:sph_surfacetension1}
\vectorbold{F}_{s \kappa,i} = -V_i \alpha_i \kappa_i \vectorbold{n}_i^{lg} \delta_i^{lg}.
\end{equation}
As proposed by Morris~\cite{Morris2000}, the curvature $\kappa_i$ is discretized according to
\begin{equation} \label{eq:sph_curvature}
\kappa_i = (\div{\vectorbold{n}^{lg}})_i = -\frac{\sum_{j} N_i N_j V_{j} \vectorbold{n}^{lg}_{ij} \pdv{W}{r_{ij}} \vectorbold{e}_{ij}}{\sum_{j} N_i N_j V_{j} W_{ij}} 
\quad \text{with} \quad 
N_k = \left\{\begin{array}{ll}
                                 1 , & \text{if}  || \grad{c}^{lg}_k || > \epsilon\\
                                0, & \text{else}
                    \end{array}\right.
\end{equation}
with $\vectorbold{n}^{lg}_{ij}=\vectorbold{n}^{lg}_{i}-\vectorbold{n}^{lg}_{j}$. Here, $\epsilon \ll 1$ is a user-defined tolerance applied to avoid contributions from particles far away from the interface with erroneous normal vectors~\cite{Morris2000}. The discrete Marangoni forces read
\begin{equation} \label{eq:sph_surfacetension2}
\vectorbold{F}_{s m,i} = V_i \underbrace{\left(\vectorbold{I} -  \vectorbold{n}^{lg}_i \otimes \vectorbold{n}^{lg}_i \right) (\grad{T})_i}_{\nabla_T {T}_i} \, \alpha_i' \delta^{lg}_i
\end{equation}
where the operator $\nabla_T$ represents the projection of the nabla operator into the interface tangential plane. Liquid-gas interfaces typically go along with considerably jumps in the mechanical and thermal constitutive parameters. It will be demonstrated in Section~\ref{sec:numex_tantempgrad} that a proper 
discretization of the temperature gradient $(\grad{T})_i$ is of upmost importance to represent the temperature field, and its inherent kink across the interface, with sufficient accuracy. Thereto, three 
different SPH gradient approximations typically applied in the literature (see also~\cite{Tong2014, Russell2018} in the context of tangential (Marangoni) surface tension forces) shall be considered in the following. The first variant is given by the standard SPH gradient approximation according to~\eqref{sph_general_gradient}:
\begin{equation} \label{eq:sph_tempgrad1}
(\grad{ T})_i \approx \sum_{j} V_{j} T_{j} \pdv{W}{r_{ij}} \vectorbold{e}_{ij} \,.
\end{equation}
The second variant is given by a symmetric gradient approximation as typically applied to gradients (e.g. of the pressure field)  in the momentum equation to guarantee for conservation of momentum~\cite{Adami2012, Adami2013}:
\begin{equation} \label{eq:sph_tempgrad2}
(\grad{ T})_i \approx \frac{1}{V_{i}} \sum_{j} (V_{i}^2+V_{j}^2) \frac{T_{i}+T_{j}}{2} \pdv{W}{r_{ij}} \vectorbold{e}_{ij} \,.
\end{equation}
The third variant is given by an anti-symmetric gradient approximation as typically applied to velocity gradients in the continuity equation to guarantee for zero-order consistency~\cite{Monaghan2005}:
\begin{equation} \label{eq:sph_tempgrad3}
(\grad{ T})_i \approx \sum_{j} V_{j} (T_{j}- T_{i}) \pdv{W}{r_{ij}} \vectorbold{e}_{ij} \,.
\end{equation}
In Section~\ref{sec:numex_tantempgrad} it will be demonstrated that only the third variant~\eqref{eq:sph_tempgrad3} leads to reasonable approximations and small discretization errors for the tangential projection of the temperature gradient. Consequently, this variant is applied in all the remaining examples in Section~\ref{sec:numex}. Moreover, it will be derived why this formulation results in a good approximation quality for the tangential projection of the temperature gradient - as required in~\eqref{eq:sph_surfacetension2} - even though it is not suitable to represent the total temperature gradient.

\subsubsection{Wetting forces} \label{subsec:nummeth_sph_wetting}

As indicated by Breinlinger et al.~\cite{Breinlinger2013}, a direct SPH discretization of~\eqref{eq:fluid_wetting} is typically not preferable due to the erroneous representation of the liquid-gas interface normal 
vector $\vectorbold{n}^{lg}$ close to the triple line solid-liquid-gas, which can be traced back to a lack of liquid and gas particle support in this region. Therefore, we follow an alternative strategy 
proposed in~\cite{Breinlinger2013} prescribing this normal vector in the triple line region on the basis of the equilibrium wetting angle $\theta_0$ according to:
\begin{equation} \label{sph:fluid_wetting1}
\hat{\vectorbold{n}}^{lg}_i = \vectorbold{t}^{sf}_i \sin \theta_0 - \vectorbold{n}^{sf}_i \cos \theta_0,
\end{equation}
where $\vectorbold{n}^{sf}_i$ is the normal vector of the solid-fluid interface according to~\eqref{eq:sph_colorfield}-\eqref{eq:sph_normalanddelta} and $\vectorbold{t}^{sf}_i$ is given by~\eqref{eq:fluid_wetting2} evaluated for particle $i$. In order to prescribe the value $\hat{\vectorbold{n}}^{lg}_i$ for the liquid-gas interface normal vector $\vectorbold{n}^{lg}_i$ in the triple point region and to have a smooth transition from the interface 
region (with prescribed normal) to the interior domain (with solution-dependent normal), the following correction scheme is employed:
\begin{equation} \label{sph:fluid_wetting3}
\vectorbold{n}^{lg}_i = \frac{f_{\vectorbold{n},i} (\grad{c^{lg}}_i / || \grad{c^{lg}}_i ||) + (1-f_{\vectorbold{n},i}) \hat{\vectorbold{n}}^{lg}_i }{|| f_{\vectorbold{n},i} (\grad{c^{lg}}_i / || \grad{c^{lg}}_i ||) + (1-f_{\vectorbold{n},i}) \hat{\vectorbold{n}}^{lg}_i  ||} \quad 
\text{with} \quad 
f_{\vectorbold{n},i}  = \left\{\begin{array}{ll}
                                 0 , & \text{if} \,\, d_{w,i} < 0\\
                                 \frac{d_{w,i}}{d_{max}} , & \text{if} \,\, 0 \leq d_{w,i} \leq d_{max}\\
                                1 , & \text{if} \,\, d_{w,i} >  d_{max}.
                    \end{array}\right.
\end{equation}
In the present work, $d_{max}=h$ has been chosen as the kernel smoothing length (which differs from~\cite{Breinlinger2013}, where the kernel support radius has been chosen) and the distance function $d_{w,i}$ is defined as the distance of fluid particle $i$ from the closest wall particle $j$ minus the initial particle spacing $h$ according to $d_{w,i} = \min ( (\vectorbold{r}_i-\vectorbold{r}_j) \cdot \vectorbold{n}^{sf}_i) - h$. From~\eqref{sph:fluid_wetting3} it becomes clear that for particles closer to the wall than $h$ the 
interface normal is prescribed as $\hat{\vectorbold{n}}^{lg}_i$, while for particles with wall distance larger than $2h$ the conventional (solution-dependent) calculation of the interface normal via the color field 
gradient is applied.

\subsubsection{Recoil pressure forces} \label{subsec:nummeth_sph_recoil}

The discrete version of the evaporation-induced recoil pressure forces occurring in~\eqref{eq:sph_momentum} is given by
\begin{equation} \label{eq:sph_recoil}
\vectorbold{F}_{v,i} = -V_i p_{v,i} \vectorbold{n}_i^{lg} \delta_i^{lg},
\end{equation}
where $p_{v,i}$ is the recoil pressure according to~\eqref{eq:fluid_recoil} evaluated for particle $i$.

\subsubsection{Viscous interface forces} \label{subsec:nummeth_sph_dissipation}

Monaghan and Gingold~\cite{Monaghan1983} proposed a stabilization term, denoted as artificial viscosity, as additional contribution to the momentum equation to reduce 
spurious flow oscillations in the numerical solution. In this work we propose to employ this stabilization term selectively only at the solid-liquid and liquid-gas interface 
to avoid oscillations originating from the phase transition solid-liquid and from the high liquid-gas interface forces typical for metal AM melt pool hydrodynamics. Moreover, it will 
be shown that the resulting interface viscosity contributions can also be motivated from a physical point of view. In its general form the discrete version of these dissipative 
interface forces is given by
\begin{equation} \label{eq:sph_dissipation}
\vectorbold{F}_{d,i} = - m_i {\zeta_i} \sum_{j} m_j \bar{h}_{ij} \bar{c}_{ij} \frac{{\vectorbold{u}_{ij}} \cdot \vectorbold{r}_{ij}}{\bar{\rho}_{ij} (r_{ij}^2 + \epsilon h_{ij}^2 )} \pdv{W}{r_{ij}},
\end{equation}
with the inter-particle averaged particle spacing $\bar{h}_{ij}=(h_i+h_j)/2$, speed of sound $\bar{c}_{ij}=(c_i+c_j)/2$ and density $\bar{\rho}_{ij}=(\rho_{i}+\rho_{j})/2$ as well as relative velocity {$\vectorbold{u}_{ij}=\vectorbold{u}_{i}-\vectorbold{u}_{j}$} and distance ${r}_{ij}=||\vectorbold{r}_{i}-\vectorbold{r}_{j}||$. The constant $\epsilon \ll 1$ is introduced to ensure a non-zero denominator. The viscosity factor is split into two contributions ${\zeta_i}={\zeta}^{lg}_i+{\zeta}^{sl}_i$. The first one, acting on (the liquid side of) the liquid-gas interface is given by:
\begin{equation} \label{eq:sph_dissipation_lg}
{\zeta}^{lg}_i = {\zeta}^{lg}_0  \delta^{lg}_i.
\end{equation}
Spurious interface flows are a well-known problem of continuum surface force (CSF) formulations~\cite{Breinlinger2013,Brackbill1996}. Due to the high magnitude of surface tension and recoil pressure 
forces this undesirable effect is in particular critical for metal AM melt pool problems. As demonstrated in Section~\ref{sec:numex_droplet_oscillation}, the introduction of an additional viscous term acting exactly (and only) at the origin of these spurious interface flows, i.e. selectively at the liquid-gas interface, enables to effectively reduce this numerical artifact {without introducing additional dissipation of physically relevant flow characteristics in the interior fluid domain}. Since these spurious interface flows are known to decrease with increasing discretization resolution~\cite{Brackbill1996}, we recommend to scale the viscosity factor ${\zeta}^{lg}_0$ with the smoothing length $h_i$. With this strategy the maximal viscous forces acting on interface particles is discretization-independent (as the maximal magnitude of $\delta^{lg}_i$ scales with $1/h_i$), while the overall influence of the viscous interface forces on the global system behavior decreases (as the interface thickness decreases with $h_i$). Similar to slope limiting techniques~\cite{Berger2005}, the numerical scheme could be further refined by applying this interface stabilization term only at locations with extremely high (or fast changing) velocity gradients or in case of metal AM melt pool simulations only at locations with very high temperatures (and thus very high recoil pressure forces). The viscosity factor on the solid-liquid interface is defined as:
\begin{equation} \label{eq:sph_dissipation_sl}
{\zeta}^{sl}_i = {\zeta}^{sl}_0 f_{{\zeta}^{sl}}(T_i) \quad \text{with} \quad
f_{{\zeta}^{sl}}(T_i)  = \left\{\begin{array}{ll}
                                 0 , & \text{if} \,\, T_i > T_{max}\\
                                 \frac{T_i-T_{max}}{T_m-T_{max}} , & \text{if} \,\, T_{max} \geq T_i \geq T_{m}\\
                                1 , & \text{if} \,\, T_i <  T_{m},
                    \end{array}\right.
\end{equation}
where $T_{m}$ is the melt temperature and $T_{max}$ a temperature chosen slightly above the melt temperature. Thus, the viscous force~\eqref{eq:sph_dissipation_sl} only 
acts on particles with temperatures close to the melting point and thus effectively damps potential instabilities arising from the jump of material parameters and state variables of a particle when undergoing the phase change solid-liquid.  Due to the no-slip condition already applied to the fluid velocity field at this interface, the influence of this additional viscous force on the global system behavior is small as long as $T_{max}$ is chosen sufficiently close to $T_{m}$. As demonstrated in~\cite{Monaghan2006}, the action of the viscous force~\eqref{eq:sph_dissipation} can be associated with an equivalent physical viscous force with effective kinematic viscosity $\nu_i = 0.5 {{\zeta}_i} \bar{h}_{ij} \bar{c}_{ij} / (d+2)$, where $d=2,3$ is the spatial dimension. Thus, besides their stabilizing effect, the contributions~\eqref{eq:sph_dissipation_lg} and~\eqref{eq:sph_dissipation_sl} can also be interpreted from a physical point of view. For example,~\eqref{eq:sph_dissipation_lg} can be thought of as part of a non-conservative surface tension formulation with interface viscosity~\cite{Gounley2016}. In particular,~\eqref{eq:sph_dissipation_sl} can be interpreted as a physical model for the gradual phase transition of alloys between solidus and liquidus temperature {such that the viscosity decreases with increasing temperature during melting.}

\subsection{Energy equation} \label{subsec:nummeth_sph_energy}

The discrete version of the energy equation~\eqref{eq:fluid_energy} has the following general form:
\begin{equation} \label{eq:sph_energy}
\dv{T_i}{t} = \frac{1}{c_{p,i} \rho_i} [- (\div \vectorbold{q})_i + \tilde{s}^{lg}_{v,i} + \tilde{s}^{lg}_{l,i}]
\end{equation}
For the discretization of the conductive term, we follow a formulation proposed by Cleary and Monaghan~\cite{Cleary1999}, which is especially suited for problems 
involving a jump of the thermal conductivity $k$ across an interface:
\begin{equation} \label{eq:sph_divq}
(\div \vectorbold{q})_i = \sum_{j} \frac{m_{j} 4k_ik_j (T_j-T_i)}{\rho_j (k_i+k_j) r_{ij}} \pdv{W}{r_{ij}}
\end{equation}
The discrete versions of the laser beam source term $\tilde{s}^{lg}_{l,i}$ and the evaporation-induced heat loss term $\tilde{s}^{lg}_{v,i}$ result directly from 
evaluating~\eqref{eq:fluid_heatsource} and~\eqref{eq:fluid_evaporation} for the discrete particle $i$.

\subsection{Equation of state} \label{subsec:nummeth_sph_eos}
The discrete version of the equation of state results from evaluating~\eqref{eq:fluid_eos} for the discrete particle $i$.
\subsection{Boundary conditions} \label{subsec:nummeth_sph_bdrycond}
\paragraph{Rigid wall boundary conditions}
Following the approach of Adami et al.~\cite{Adami2012} rigid wall boundary conditions are modeled using fixed boundary particles with quantities extrapolated from the fluid field based on a local force balance. The same approach is used to model the mechanical interaction between fluid particles and the solid phase. For more details the interested reader is referred to the aforementioned literature.
\paragraph{Periodic boundary conditions}
Imposing a periodic boundary condition in a specific spatial direction allows for particle interaction evaluation across the lower and upper domain border. Moreover, particles leaving the domain on one side are re-injecting on the opposite side.
\section{Time integration scheme} \label{sec:nummeth_sph_timint}
The momentum equation~\eqref{eq:sph_momentum} is integrated in time applying an explicit velocity-Verlet time integration scheme in kick-drift-kick form, also denoted as leapfrog scheme, as proposed by Monaghan~\cite{Monaghan2005}. In the absence of dissipative effects the velocity-Verlet scheme is of second order accuracy and reversible in time~\cite{Monaghan2005}. In a first kick-step the particle accelerations $\vectorbold{a}_{i}^{n} = \qty(\dv*{\vectorbold{u}_{i}}{t})^{n}$ determined in the previous time step~$n$ are used to compute intermediate particle velocities at~$n+1/2$
\begin{equation}
\vectorbold{u}_{i}^{n+1/2} = \vectorbold{u}_{i}^{n} + \frac{\Delta{}t}{2} \, \vectorbold{a}_{i}^{n} \, ,
\end{equation}
where~$\Delta{}t$ is the time step size, before the particle positions at~$n+1$ are updated in a drift-step
\begin{equation}
\vectorbold{r}_{i}^{n+1} = \vectorbold{r}_{i}^{n} + \Delta{}t\vectorbold{u}_{i}^{n+1/2} \, .
\end{equation}
With the particle positions~$\vectorbold{r}_{i}^{n+1}$ the densities~$\rho_{i}^{n+1}$ are determined on the basis of~\eqref{eq:sph_densum}. Based on the temperatures~${T}_{i}^{n}$ 
as well as the updated particle positions~$\vectorbold{r}_{i}^{n+1}$ and densities~$\rho_{i}^{n+1}$ the temperature rate $\left(d T_i / d t\right)^{n+1}$ is calculated 
on the basis of~\eqref{eq:sph_energy} and the temperature is updated according to:
\begin{equation}
{T}_{i}^{n+1} = {T}_{i}^{n} + \Delta{}t \left(\frac{d T_i}{d t}\right)^{n+1} \, .
\end{equation}
Using the updated particle temperatures~${T}_{i}^{n+1}$, positions~$\vectorbold{r}_{i}^{n+1}$ and densities~$\rho_{i}^{n+1}$ as well as the intermediate velocities $\vectorbold{u}_{i}^{n+1/2}$ the accelerations~$\vectorbold{a}_{i}^{n+1}$ are calculated from~\eqref{eq:sph_momentum}. In a final kick-step the particle velocities at time step~$n+1$ are determined via
\begin{equation}
\vectorbold{u}_{i}^{n+1} = \vectorbold{u}_{i}^{n+1/2} + \frac{\Delta{}t}{2} \, \vectorbold{a}_{i}^{n+1} \, .
\end{equation}
To maintain stability of the time integration scheme the time step size~$\Delta{}t$ is restricted by the Courant-Friedrichs-Lewy (CFL) condition, the viscous condition, the body force condition, the surface tension condition, and the conductivity-condition refer to~\cite{Adami2010, Morris1997, Adami2013, Cleary1998} for more details,
\begin{equation} \label{eq:sph_timestepcond}
\Delta{}t \leq \min\qty{ 0.25\frac{h}{c+\qty|\vectorbold{u}_{max}|}, \quad 0.125\frac{h^{2}}{\nu}, \quad 0.25\sqrt{\frac{h}{\qty|\vectorbold{b}_{max}|}} ,
\quad 0.25\sqrt{\frac{\rho h^3}{2 \pi \alpha}} , \quad 0.125 \frac{\rho c_p h^2}{k} } \, ,
\end{equation}
with kinematic viscosity $\nu=\eta/\rho$, maximum fluid velocity~$\vectorbold{u}_{max}$ and maximum body force~$\vectorbold{b}_{max}$.\\
\section{Comparison of different temperature gradient approximations} \label{sec:numex_tantempgrad}

\begin{figure}[htbp]
\centering
\subfigure [Temperature field ranging from $1700$ (blue) to $2700$ (red).]
{
\includegraphics[width=0.47\textwidth]{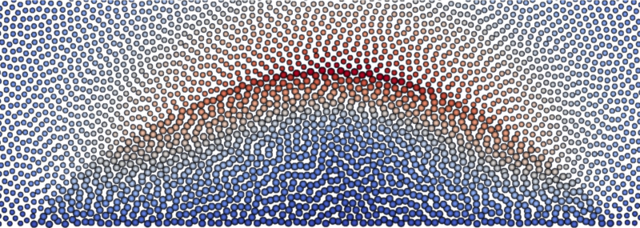}%
\label{fig:example1_tempandcolorfield_a}
}
\hspace{0.01\textwidth}
\subfigure [Total color field ranging from $0.99$ (blue) to $1.01$ (red).]
{
\includegraphics[width=0.47\textwidth]{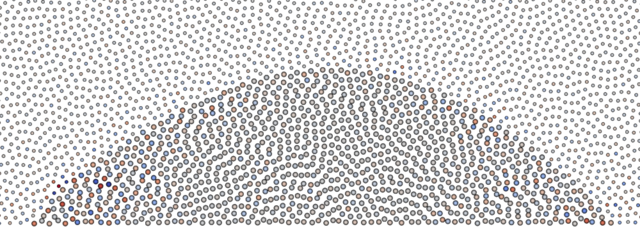}%
\label{fig:example1_tempandcolorfield_b}
}
\caption{Laser heating of liquid drop (large particles) on solid substrate (not displayed) surrounded by gas (small particles).}
\label{fig:example1_tempandcolorfield}
\end{figure}

\begin{figure}[htbp]
\centering
\subfigure [Variants \textit{CSPM}~\cite{Chen2001} vs. \textit{CSPH}~\cite{Bonet1999} (shaded).]
{
\includegraphics[width=0.45\textwidth]{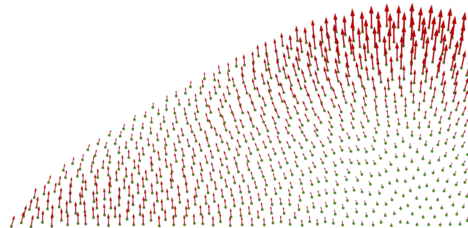}%
\label{fig:example1_tempgrad_a}
}
\hspace{0.01\textwidth}
\subfigure [Variants \textit{Standard}~\eqref{eq:sph_tempgrad1} vs. \textit{CSPH}~\cite{Bonet1999} (shaded).]
{
\includegraphics[width=0.45\textwidth]{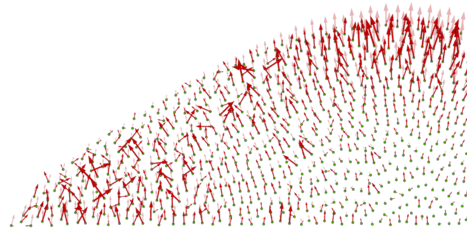}%
\label{fig:example1_tempgrad_b}
}
\subfigure [Variants \textit{Symmetric}~\eqref{eq:sph_tempgrad2} vs. \textit{CSPH}~\cite{Bonet1999} (shaded).]
{
\includegraphics[width=0.45\textwidth]{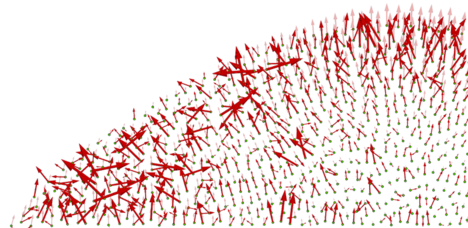}%
\label{fig:example1_tempgrad_c}
}
\hspace{0.01\textwidth}
\subfigure [Variants \textit{Asymmetric}~\eqref{eq:sph_tempgrad3} vs. \textit{CSPH}~\cite{Bonet1999} (shaded).]
{
\includegraphics[width=0.45\textwidth]{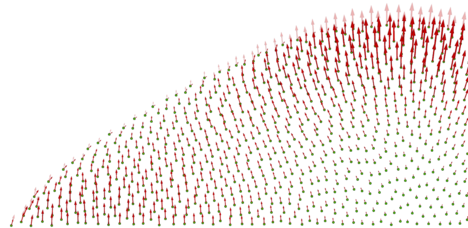}%
\label{fig:example1_tempgrad_d}
}
\caption{Laser heating of liquid drop: different approximations for total temperature gradient. Displayed is half of the droplet. The total temperature gradient is visualized by bright (investigated solution) and shaded (reference solution) red arrows.}
\label{fig:example1_tempgrad}
\end{figure}

\begin{figure}[htbp]
\centering
\subfigure [Variants \textit{CSPM}~\cite{Chen2001} vs. \textit{CSPH}~\cite{Bonet1999} (shaded).]
{
\includegraphics[width=0.45\textwidth]{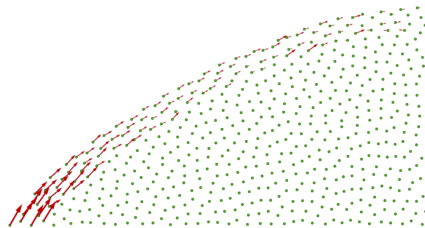}%
\label{fig:example1_fig:example1_tempgradtan_a}
}
\hspace{0.01\textwidth}
\subfigure [Variants \textit{Standard}~\eqref{eq:sph_tempgrad1} vs. \textit{CSPH}~\cite{Bonet1999} (shaded).]
{
\includegraphics[width=0.45\textwidth]{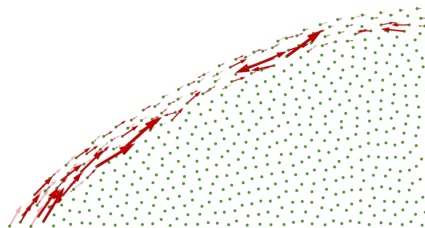}%
\label{fig:example1_fig:example1_tempgradtan_b}
}
\subfigure [Variants \textit{Symmetric}~\eqref{eq:sph_tempgrad2} vs. \textit{CSPH}~\cite{Bonet1999} (shaded).]
{
\includegraphics[width=0.45\textwidth]{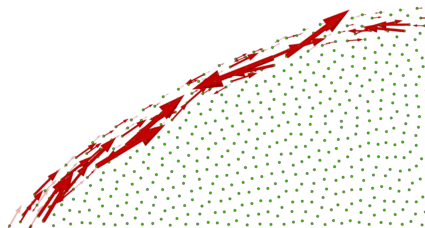}%
\label{fig:example1_fig:example1_tempgradtan_c}
}
\hspace{0.01\textwidth}
\subfigure [Variants \textit{Asymmetric}~\eqref{eq:sph_tempgrad3} vs. \textit{CSPH}~\cite{Bonet1999} (shaded).]
{
\includegraphics[width=0.45\textwidth]{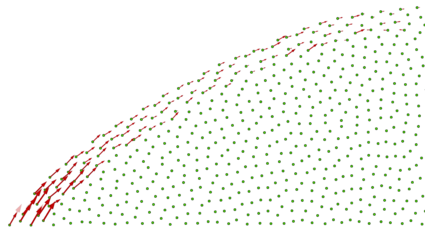}%
\label{fig:example1_fig:example1_tempgradtan_d}
}
\caption{Laser heating of liquid drop: different approximations for tangential temperature gradient. Displayed is half of the droplet. The tangential gradient is visualized by bright (investigated solution) and shaded (reference solution) red arrows.}
\label{fig:example1_fig:example1_tempgradtan}
\end{figure}

In this section, the approximation quality of the different temperature gradient discretizations~\eqref{eq:sph_tempgrad1}-\eqref{eq:sph_tempgrad3} will be investigated. For this 
purpose the temperature field in the liquid and gas phase of a liquid drop resting on a solid substrate and surrounded by a gas atmosphere will be considered at a representative time step (see Figure~\ref{fig:example1_tempandcolorfield_a}). A detailed description of the problem setup is given in Section~\ref{sec:numex_heateddrop}, where the 
full thermo-hydrodynamic interaction within this problem is studied. In Figure~\ref{fig:example1_tempandcolorfield_b}, the total color field $\hat{c}_i:=\sum_{j} V_{j} W_{ij}$, which considers contributions from all possible types (i.e. gas, liquid and solid phase) of neighbor particles $j$, is displayed for the considered droplet example. The fact that the color field is close to one throughout the entire domain suggests that all particles have full support and, thus, \textit{two-sided} gradient approximations such as~\eqref{eq:sph_tempgrad1}-\eqref{eq:sph_tempgrad3} 
might be applicable in general.
It is typically argued that \textit{one-sided} SPH gradient approximations such as the corrective smoothed particle method (CSPM) by Chen et al.~\cite{Chen2001} or the Corrected SPH (CSPH) scheme by Bonet and Lok~\cite{Bonet1999}, which allow for exact representation of first-order gradients even in boundary or interface regions with incomplete particle support, are required to 
accurately capture a kink in the temperature field resulting from the jump in the thermal conductivity $k$ at fluid-gas interfaces~\cite{Tong2014}. Therefore, the first-order consistent schemes \textit{CSPM} as well as \textit{CSPH} will be considered as reference solutions in this section and compared to the temperature gradient discretizations according to~\eqref{eq:sph_tempgrad1} (variant \textit{Standard}),~\eqref{eq:sph_tempgrad2} (variant \textit{Symmetric}) and~\eqref{eq:sph_tempgrad3} (variant \textit{Asymmetric}). In Figure~\ref{fig:example1_tempgrad}, the temperature gradients 
associated with the temperature field~\ref{fig:example1_tempandcolorfield_a} are displayed for the variants \textit{CSPM}, \textit{Standard},  \textit{Symmetric} and  \textit{Asymmetric} and compared 
to the variant \textit{CSPH} (shaded). As expected, there is no visual difference of the two \textit{one-sided} gradient approximations (variants \textit{CSPM} and \textit{CSPH} in Figure~\ref{fig:example1_tempgrad_a}). Indeed, the results of these two formulations are identical up to machine precision. However, the non-smooth temperature field across the interface leads to deviations for the other three variants in this region (Figures~\ref{fig:example1_tempgrad_b}-~\ref{fig:example1_tempgrad_d}). While these visible deviations vanish in the interior of the drop for the variant \textit{Asymmetric}, the variants \textit{Standard} and \textit{Symmetric} show large deviations also in this domain, which seems to contradict first intuition. However, this observation can be explained by the fact that the asymmetric formulation exactly filters out constant temperature contributions (exactly vanishing gradient for constant fields), while large constant temperature contributions lead to considerable discretization errors for the variants \textit{Standard} and \textit{Symmetric} due to the lack of zero-order consistency, i.e. $\int_{\Omega} \nabla W\qty(\qty| \vectorbold{r} - \vectorbold{r}' |, h) \dd{\vectorbold{r}'} = 0$ but $\sum_{j} V_{j} \pdv{W}{r_{ij}} \vectorbold{e}_{ij}\neq 0$. The avoidance of large constant contributions is also the reason why the reference pressure is typically set to zero in SPH formulations where the pressure gradient is approximated by the momentum-conserving variant \textit{Symmetric}~\cite{Morris1997}.
When considering the tangential projection of the temperature gradient $\nabla_T {T}_i$ as defined in~\eqref{eq:sph_surfacetension2} and displayed in Figure~\ref{fig:example1_fig:example1_tempgradtan}, the observations made above for the variants \textit{CSPH}, \textit{CSPM}, \textit{Standard},  and \textit{Symmetric} can be confirmed. Specifically the latter two variants are not suitable to represent the 
tangential temperature gradient. In contrast, the variant \textit{Asymmetric} represents the tangential temperature gradient very well without visual difference to the variants \textit{CSPH} and \textit{CSPM} - except for a small deviation at the triple point, which can be expected due to the non-smooth interface line in this region. Therefore, the variant \textit{Asymmetric} according to~\eqref{eq:sph_tempgrad3} will be employed - and favored over the computationally more involved \textit{CSPH} and \textit{CSPM} schemes - in all the remaining numerical examples in this work. In the remainder of this section, a brief analytical explanation will be given on why the \textit{two-sided} gradient approximation~\eqref{eq:sph_tempgrad3} is capable of correctly representing the tangential gradient projection of a temperature field with a kink at the interface.
Let's assume a particle $i$ close to the liquid-gas interface with closest / orthogonal projection point $c$ onto this interface. Furthermore, a coordinate system with axes 
$\vectorbold{t}_1$ and $\vectorbold{t}_2$ tangential to the interface and axis $\vectorbold{n}$ normal to the interface is defined in  point $c$. The position of a material particle expressed in this system
shall be denoted as $\tilde{\vectorbold{x}}$, and a quantity evaluated at position $c$ shall be marked by a subscript $(...)_c$. Moreover, subscripts $(...)_n$ and $(...)_t$ represent projections of vectors in interface normal and tangential direction. The first-order Taylor series expansion of the temperature field in the neighborhood of point $c$ is given as:
\begin{equation} \label{eq:examp1_tempfield}
T(\tilde{\vectorbold{x}}) \approx T_c + \grad{T}_c \cdot \tilde{\vectorbold{x}} = T_c + \boldsymbol{\nabla}_t T_c \cdot \tilde{\vectorbold{x}}_t + \boldsymbol{\nabla}_n T_c \cdot \tilde{\vectorbold{x}}_n 
\quad \text{with} \quad 
\boldsymbol{\nabla}_n T_c = \left\{\begin{array}{ll}
                                 \boldsymbol{\nabla}_n T_c^+ , & \text{if} \,\, \tilde{\vectorbold{x}}_n \cdot \vectorbold{n} > 0\\
                                \boldsymbol{\nabla}_n T_c^-, & \text{else}.
                    \end{array}\right.
\end{equation}
Note that~\eqref{eq:examp1_tempfield} accounts for a jump of the temperature gradient in normal direction as typical for liquid-gas interfaces. According to~\eqref{sph_general_gradient} the projection of the temperature gradient in the tangential directions $\vectorbold{t}_k$ reads:
\begin{equation}
\label{eq:examp1_gradient1}
\vectorbold{t}_k \cdot \grad{T}\qty(\vectorbold{r}_i) \approx \vectorbold{t}_k \cdot \int_{\Omega_i} [T_c + \boldsymbol{\nabla}_t T_c \cdot \tilde{\vectorbold{x}}_t + \boldsymbol{\nabla}_n T_c \cdot \tilde{\vectorbold{x}}_n] \, \grad{W} \dd{\tilde{\vectorbold{x}}} \quad \text{for} \quad k=1,2.
\end{equation}
In the following, only the third of the three summands will be considered. For simplicity, the integral is reformulated in spherical coordinates $(r,\alpha, \beta)$ using $\tilde{\vectorbold{x}}= r \vectorbold{e}$ with $\vectorbold{e}=\sin \beta \cos \alpha \vectorbold{t}_1 + \sin \beta \sin \alpha \vectorbold{t}_2 + \cos \beta \vectorbold{n}$:
\begin{equation}
\label{eq:examp1_gradient2}
\begin{split}
& \vectorbold{t}_k \cdot \int_{\Omega_i} \boldsymbol{\nabla}_n T_c \cdot \tilde{\vectorbold{x}}_n \, \grad{W} \dd{\tilde{\vectorbold{x}}} \\
= & \boldsymbol{\nabla}_n T_c \cdot \int_{r=0}^{r_c} \int_{\beta=0}^{\pi} \int_{\alpha=0}^{2\pi}  \underbrace{r \cos \beta \vectorbold{n}}_{\tilde{\vectorbold{x}}_n} \underbrace{\pdv*{W}{r} \vectorbold{e}}_{\grad{W}} \cdot \vectorbold{t}_k \underbrace{r^2 \sin \beta  \dd{\alpha} \dd{\beta} \dd{r}}_{\dd{\tilde{\vectorbold{x}}}} \\
= & \boldsymbol{\nabla}_n T_c \! \cdot \! \vectorbold{n} \int_{r=0}^{r_c} r^3 \pdv*{W}{r} \int_{\beta=0}^{\pi} \cos \beta \sin^2 \beta 
\underbrace{\int_{\alpha=0}^{2\pi} [\cos \alpha \vectorbold{t}_1 \! \cdot \! \vectorbold{t}_k  + \sin \alpha \vectorbold{t}_2  \! \cdot \! \vectorbold{t}_k]   \dd{\alpha}}_{=0} \dd{\beta} \dd{r} = 0.
\end{split}
\end{equation}
From the second to the third line in~\eqref{eq:examp1_gradient2} use have been made of the spherical symmetry of the kernel $W$, i.e. $\pdv*{W}{r}$ is independent of $\alpha$ and $\beta$. {It is emphasized that only for spherically symmetric kernels the tangential projection of the SPH discretization of a gradient contains only information of the tangential component of the space-continuous gradient, i.e. it is independent of the evolution of the underlying field in normal direction and the third summand in~\eqref{eq:examp1_gradient1} vanishes.} Thus, it can be concluded that standard (two-sided) gradient discretization approaches can be applied since the gradient jump in normal direction does not enter the SPH formulation for the tangential gradient discretization. While this derivation has been made for the \textit{standard} temperature gradient approximation~\eqref{eq:sph_tempgrad1}, the results are equally valid for the \textit{symmetric}~\eqref{eq:sph_tempgrad2} and \textit{asymmetric}~\eqref{eq:sph_tempgrad3} gradient approximation since the main difference is a constant temperature value $T_i$ added to or subtracted from the standard gradient approximation.

\section{Numerical examples} \label{sec:numex}
\subsection{Liquid droplet in surrounding fluid} \label{sec:numex_migratingbubble}

To verify the proposed formulation for temperature-dependent surface tension, cf. Section~\ref{subsec:nummeth_sph_surface tension}, the migration of a liquid droplet as proposed by Ma and Bothe~\cite{Ma2011} is considered. While a finite volume scheme is employed in~\cite{Ma2011}, the same problem has been studied by Tong and Browne~\cite{Tong2014} using an incompressible SPH formulation. The problem consists of a circular droplet (radius $a = 1.44$) of fluid~1 (density $\rho^1=0.25$, dynamic viscosity $\mu^1 = 12.0$, thermal conductivity $k^1=1.2 \times 10^{3}$, heat capacity $c_p^{1} = 50.0$) that is initially resting at the center of a quadratic domain (side length $4a$) filled with fluid~2 ($\rho^2=2\rho^1$, $\mu^2 = 2 \mu^1$, $k^2=2 k^1$, $c_p^{2} = 2 c^{1}$). The temperature-dependent surface tension is defined via $\alpha_0=1.0 \cdot 10^{4}$, $\alpha'_0=2.0 \cdot 10^{3}$ and $T_{\alpha_0}=290$. All properties are given in the units $mm$, $mg$, $s$, and $K$. On the left and right side of the quadratic domain periodic boundary conditions are applied. At the top and bottom of the quadratic domain no-slip boundary conditions for the fluid field as well as prescribed temperatures $\hat{T}^{1}$ (bottom) and $\hat{T}^{2}$ (top) are applied via three layers of boundary particles according to~\cite{Adami2012}. The 2D domain is discretized by a total of 4096 particles (812 particles for fluid 1, 3284 particle for fluid 2) resulting in an initial particle spacing of $\Delta x = 0.09$. Initially, the particles are at rest, i.e. $\boldsymbol{u}_0^1=\boldsymbol{u}_0^2=\boldsymbol{0}$. Like in the original works, the surface tension acts with its full magnitude during the entire simulation time, i.e. no initial ramp function is used. Moreover, the initial temperature profile $T_0$ is chosen 
as linear interpolation between  $\hat{T}^{1}$ and $\hat{T}^{2}$ leading to an initial gradient of $||\grad{T_0}||=(\hat{T}_{2}-\hat{T}_{1})/(4a)$.

\subsubsection{Static equilibrium of droplet}

As a first test case the pressure jump across the interface of the bubble in a static equilibrium configuration shall be investigated. Thereto the temperatures at the bottom and top wall are prescribed to $\hat{T}_{1}=\hat{T}_{2}=290$. The weakly compressible approach is realized with bulk moduli $K^{1} = 78.125 \cdot 10^{3}$ and $K_{2} = 156.25 \cdot 10^{3}$ (artificial speed of sound $c^1=c^2 \approx 562.5$). The time step size has been chosen to $\Delta t = 4.0 \cdot 10^{-5}$. The analytical solution for this problem setup is a constant temperature field at $T=290$ with the circular droplet of radius $a$ resting at the center of the quadratic domain. While the Marangoni forces vanish for this test case, due to the constant surface tension contribution of $\alpha=\alpha_0=1.0 \cdot 10^{4}$ a pressure jump of $\Delta p = \frac{\alpha_0}{a} \approx 6944$ is expected at the interface of the circular droplet. 
\begin{figure}[htbp]
\centering
\subfigure [Static equilibrium: pressure jump $\Delta p := p - p_{\infty}$ over distance $r$ from droplet center for sections horizontal (black) and diagonal (blue) through the droplet.]
{
\includegraphics[width=0.45\textwidth]{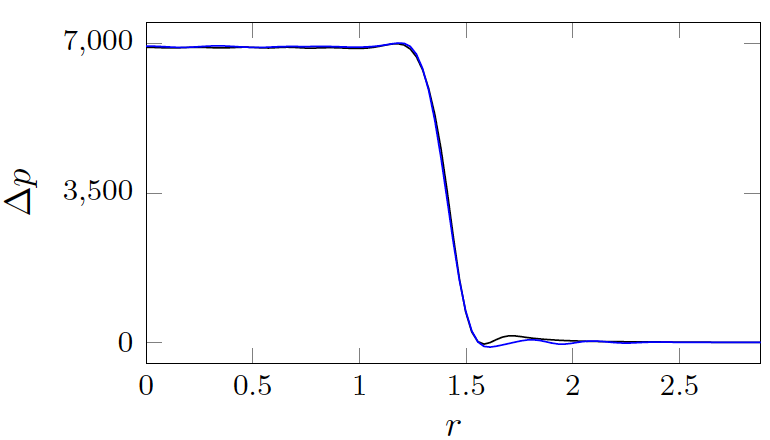}%
\label{fig:label_subfig_1}
}
\hspace{0.01\textwidth}
\subfigure [Thermo-capillary migration: dimensionless centroid velocity $U/U_{r}$ over dimensionless time $t/t_{r}$ for present formulation (solid) and reference solution ($+$ marks,~\cite{Ma2011}).]
{
\includegraphics[width=0.45\textwidth]{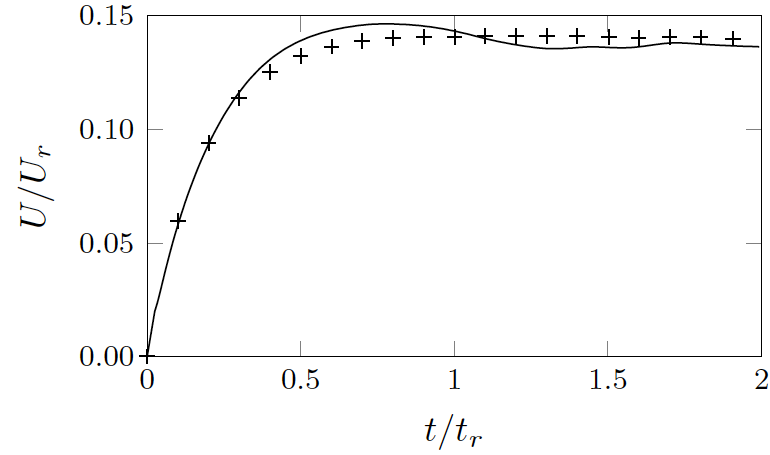}%
\label{fig:label_subfig_2}
}
\caption{Different configurations of a droplet within surrounding fluid: static equilibrium and thermo-capillary migration.}
\label{fig:example_hydrostatic_pressure}
\end{figure}
In Figure~\ref{fig:label_subfig_1} the pressure jump $\Delta p := p - p_{\infty}$, with $p_{\infty}$ representing the pressure in fluid~2 at sufficient distance from the droplet interface, is displayed over the distance $r$ from the droplet center for a horizontal (black) as well as a diagonal (45 degrees) section (blue) through the droplet center. 
Besides slight oscillations at the inner and outer fringe of the interface region, the numerical results show very good agreement with the analytical prediction of the maximal pressure jump at the droplet center given by $\Delta p \approx 6944$.

\subsubsection{Thermo-capillary migration of droplet}

As a second test case the thermo-capillary migration of the droplet will be considered. Thereto the temperatures at the bottom and top wall are prescribed to $\hat{T}_{1}=290$ and 
$\hat{T}_{2}=T_{1} + 4 a \qty|\nabla T| = 291.152$ such that the stationary initial temperature field is characterized by a constant gradient $||\grad{T_0}|| = 0.2$. For the 
simulation of this problem the bulk moduli $K^{1} = 5.0 \cdot 10^{6}$ and $K^{2} = 1.0 \cdot 10^{7}$ (artificial speed of sound $c^1=c^2 \approx 4472.1$) as well as a time step size 
of $\Delta t = 4.0 \cdot 10^{-5}$ have been chosen.
\begin{figure}[htbp]
\centering
\includegraphics[width=0.7\textwidth]{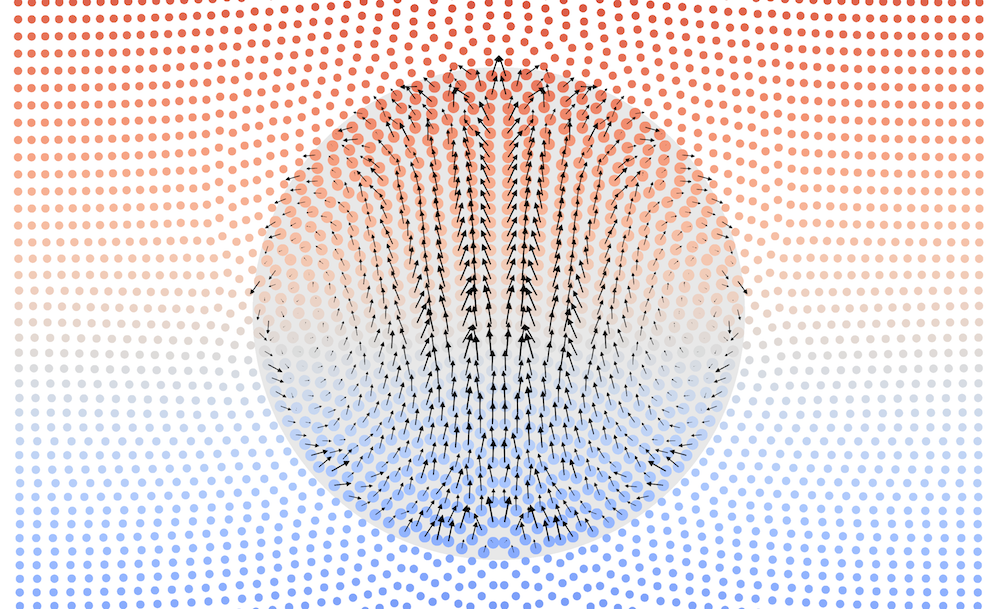}
\caption{Thermo-capillary migration of a droplet (in vertical direction only part of the domain is shown): velocity profile within droplet indicated by arrows and temperature profile of both fluid phases colored in the temperature range from $T_{1} = 290.0$ (blue) to $T_{2} = 291.152$ (red).}
\label{fig:example_migrating_droplet}
\end{figure}
In this test case, temperature-dependent surface tension in combination with the prescribed temperature gradient results in shear stresses (second term in~\eqref{eq:fluid_surfacetension}) and consequently a shear flow (Marangoni convection) at the droplet interface (see Figure~\ref{fig:example_migrating_droplet}). On the one hand, this shear flow induces an upwards migration of the droplet. On the other hand, the shear flow induced in the surrounding fluid~2 redistributes the initially linear temperature profile. In terms of dimensionless numbers the considered choice of parameters leads to a Reynolds number of $Re = \frac{\rho a U_{r}}{\mu} = 0.72$, a Marangoni number of $Ma = \frac{\rho c_{p} a U_{r}}{\lambda} = 0.72$ and a capillary number of $Ca = \frac{\mu U_{r}}{\sigma_{0}} = 0.0576$, where the characteristic velocity $U_{r} = \sigma_{t} \qty|\nabla T| \frac{a}{\mu} = 24$ has been employed. In Figure~\ref{fig:label_subfig_2} the dimensionless velocity $U/U_{r}$, with $U$ calculated via numerical differentiation of the droplet centroid position with respect to time, is plotted over the dimensionless time $t/t_{r}$ with $t_{r}= \frac{a}{U_{r}} = 0.06$. The resulting velocity evolution shows good agreement with the reference solution~\cite{Ma2011}. The slight fluctuations in the steady state regime can be traced back to the employed weakly compressible formulation, while the reference solution in~\cite{Ma2011} has been calculated on the basis of an incompressible formulation.

\subsection{Oscillation of liquid dropled surrounded by gas atmosphere} \label{sec:numex_droplet_oscillation}

\begin{figure}[htbp]
\centering
\subfigure [Configuration at t=0.]
{
\includegraphics[width=0.3\textwidth]{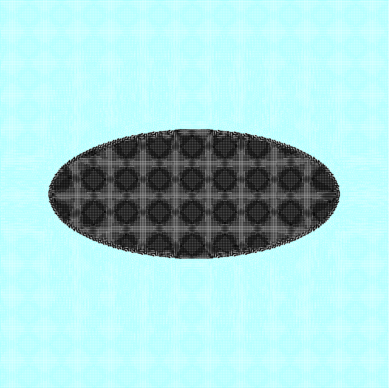}%
\label{fig:example2b_tempandcolorfield_a}
}
\hspace{0.01\textwidth}
\subfigure [Configuration at t=T/2.]
{
\includegraphics[width=0.3\textwidth]{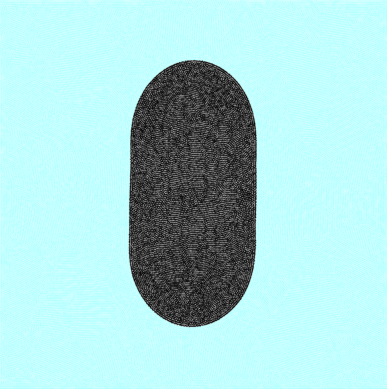}%
\label{fig:example2b_tempandcolorfield_b}
}
\hspace{0.01\textwidth}
\subfigure [Configuration at t=T.]
{
\includegraphics[width=0.3\textwidth]{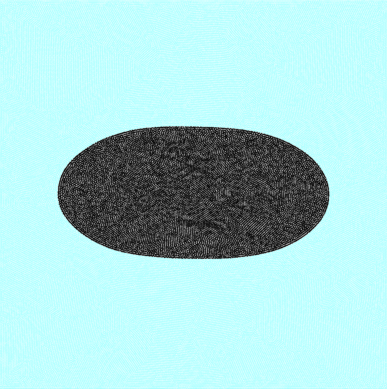}%
\label{fig:example2b_tempandcolorfield_c}
}
\caption{Oscillation of liquid dropled surrounded by gas atmosphere for reference parameter set at different times.}
\label{fig:example2b_tempandcolorfield}
\end{figure}

This example aims to investigate the influences of the viscous interface force introduced in Section~\ref{subsec:nummeth_sph_dissipation} as well as of approximated material properties of the gas phase on the solution accuracy of surface tension-dominated problems with parameter values as typical for metal AM. For this purpose the well-known example of surface tension-driven oscillations of a liquid droplet in gas atmosphere is considered~{\cite{Nugent2000, Morris2000, Hu2006, Adami2010}}. The droplet is initially resting at the center of a quadratic gas domain with side length $400 \mu m$ and has an elliptic initial shape (larger semiaxis $a=3/2R$, smaller semiaxis $b=2/3R$, where $R=100 \mu m$ is the droplet radius in the static equilibrium configuration;  see Figure~\ref{fig:example2b_tempandcolorfield_a}). The reference pressure of the weakly compressible model is set to $p_0=1.0 \cdot 10^7 N m^{-2}$ and the background pressure of the transport velocity formulation to $p_b=5p_0$ for both phases. Moreover, a time step size of $\Delta t = 10^{-6} ms$ and an initial particle spacing of $\Delta x = 5/3 \mu m$ is applied. Again, the temperature field is kept constant such that Marangoni forces vanish. The material properties of the liquid phase inside the droplet are identical to the ones used for the subsequent melt pool simulations and are given in Table~\ref{tab:material_params}. Moreover, different material properties of the gas phase will be investigated in the following. As reference parameter set the gas phase is modeled such that the density and dynamic viscosity are by a factor of $1000$ smaller as compared to the values of the liquid phase. Driven by the surface tension forces the droplet executes oscillations with period T. Figure~\ref{fig:example2b_tempandcolorfield} illustrates the droplet shapes resulting from the reference parameter set at times $t=0$, $t=T/2$ and $t=T$.
For the case of small amplitude oscillations there is an analytical solution for the resulting oscillation period~{\cite{rayleigh1879capillary}} given by $T_a=2\pi \sqrt{R^3 \rho_0^l / (6 \alpha)}$, where $\rho_0^l$ the initial density of the liquid phase. For the given set of parameters the analytical solution for the oscillation period takes on a value of $T_a=0165ms$. For the numerical solution based on the reference parameter set as illustrated in Figure~\ref{fig:example2b_tempandcolorfield} an oscillation period of 
$T_{ref}=0.179ms$ can be determined. The deviation can be explained by the fact that in the present problem setup, which is considered more representative for melt pool hydrodynamics, \textit{large} amplitude oscillations occur and the damping of the oscillation amplitude resulting from the viscosity of the liquid metal is not negligible in this scenario. Specifically, after one oscillation (i.e. at $t=T$) the initial droplet length $l(t=0)=300 \mu m$ has already decreased to $l(t=T)=280 \mu m$. To increase the computational efficiency (in terms of larger critical time step sizes) and robustness of the numerical formulation the density and dynamic viscosity of the gas phase shall be increased. This procedure is considered reasonable since the focus of this work lies on the thermo-hydrodynamics in the melt phase while an exact representation of detailed flow patterns in the gas phase is not of primary interest. When decreasing the density ratio between liquid and gas phase to $100$ and the dynamic viscosity ratio to $10$ the droplet length after one oscillation is $l(t=T)=274 \mu m$, representing a relative error of $\approx 2 \%$ with respect to the reference parameter set (density and dynamic viscosity ratio of $1000$). Since an error of $\approx 2 \%$ seems reasonable for the intended 
melt pool simulations, this parameter set ($\rho_0^l/\rho_0^g=100$, $\eta^l/\eta^g=10$) is taken as default value for the subsequent melt pool simulations. Finally, the influence of the viscous interface forces (Section~\ref{subsec:nummeth_sph_dissipation}), which should stabilize the numerical scheme without significantly 
changing the physical system behavior, will be investigated. For the discretization resolutions and recoil pressure magnitudes considered in the subsequent numerical examples values of the viscosity constant in the range of ${\zeta}_0^{lg} \sim 10^{-4}$ turned out to effectively stabilize the scheme. Considering the default parameter set ($\rho_0^l/\rho_0^g=100$, $\eta^l/\eta^g=10$) and adding additional viscous interface forces with ${\zeta}_0^{lg}=2.5 \cdot 10^{-4}$ and ${\zeta}_0^{lg}=1.0 \cdot 10^{-4}$ results in relative errors of $\approx 6 \%$ and $\approx 2 \%$ in the droplet length $l(T)$ with respect to the solution resulting from the default parameter set ($\rho_0^l/\rho_0^g=100$, $\eta^l/\eta^g=10$) with ${\zeta}_0^{lg}=0$. Again, this accuracy is considered reasonable for the intended melt pool simulations.

\subsection{Representative test cases for metal additive manufacturing applications} \label{sec:numex_am}

The remaining test cases are designed to verify the proposed formulation with respect to specific effects, conditions and material properties as typical for metal additive manufacturing processes. 
As representative material properties values close to the parameters for stainless steel at the melting point ($T_m=1700K$) according to~\cite{Khairallah2014} and~\cite{Khairallah2016} are considered. Table~\ref{tab:material_params} represents the corresponding material parameters considered for the liquid metal phase, i.e. for the melt phase. While the solid phase is modeled as rigid and immobile in this work, the thermal problem in the solid phase is solved in the same manner as for the liquid metal phase employing the same thermal material parameters.
\begin{table}[htbp]
\centering
\begin{tabular}{|p{1.5cm}|p{7.0cm}|p{3.0cm}|p{3.0cm}|} \hline
Symbol & Property & Value & Units \\ \hline
 $\rho_0$ & Initial / reference density& $7430$ & $kg \, m^{\!-3}$ \\ \hline
 $\eta$ & Dynamic viscosity& $6 \cdot 10^{-3}$ & $kg \, m^{\!-1} \, s^{\!-1}$  \\ \hline
 $\alpha_0$ & Surface tension at reference temperature & $1.8$ & $N \, m^{\!-1}$  \\ \hline
 $T_{\alpha_0}$ & Reference temperature for surface tension& $1700$ & $K$  \\ \hline
 $\alpha'_0$ & Surface tension gradient coefficient& $1.0 \cdot 10^{-3}$ & $N \, m^{\!-1} \, K^{\!-1}$  \\ \hline
 $\theta_0$ & Wetting angle & $60$ & degrees  \\ \hline
 $T_0$ & Initial temperature & $500$ & $K$  \\ \hline
 $\hat{T}$ & Temperature at Dirichlet boundaries & $500$ & $K$  \\ \hline
 $T_m$ & Melting temperature & $1700$ & $K$  \\ \hline
 $T_v$ & Evaporation / boiling temperature & $3000$ & $K$  \\ \hline
 $c_p$ & Heat capacity & $965$ & $J \, kg^{-1} \, K^{-1}$  \\ \hline
 $k$ & Thermal conductivity & $35.95$ & $ W\, m^{-1} \, K^{-1}$  \\ \hline
 $\chi_l$ & Laser absorptivity & $0.5$ & $ - $  \\ \hline
 $C_P$ & Pressure constant of recoil pressure model & $20$ & $ N \, m^{-2}$  \\ \hline
 $C_T$ & Temperature constant of recoil pressure model & $1 \cdot 10^5$ & $K$  \\ \hline
 $h_v$ & Latent heat of evaporation & $6 \cdot 10^6$ & $J \, kg^{-1}$  \\ \hline
 $T_{h,0}$ & Reference temperature for specific enthalpy & $663.731$ & $K$  \\ \hline
 $C_M$ & Constant for vapor mass flow & $1.0 \cdot 10^{-3}$ & $K \, s^2 \, m^{-2}$  \\ \hline
\end{tabular}
\caption{Representative material parameters for stainless steel taken for the liquid metal phase.}
\label{tab:material_params}
\end{table}
Note that the enthalpy reference temperature $T_{h,0}$ has been chosen such that the specific enthalpy at the melting point yields $h(T=T_m)=1 \cdot 10^6 J \, kg^{-1}$ according to~\eqref{eq:fluid_evaporation}. In order to increase the effect of recoil pressure and to make the numerical examples more challenging with respect to the robustness of the proposed formulation, the constants $C_P$ and $C_T$ of the recoil pressure model as well as the laser powers employed in Sections~\ref{sec:numex_2Dmelt_rec} and~\ref{sec:numex_3Dmelt} have been adapted compared to typical values for SLM processes. {Thus, the following examples are intended to demonstrate the robustness of the proposed SPH formulation and its general suitability for challenging application scenarios in metal AM, but not to represent the system behavior of a specific AM experiment with given process and material parameters. In our continued research work additional modeling aspects such as temperature-dependent material parameters or laser beam absorption via ray tracing, capturing e.g. multiple reflection and shading effects, will be incorporated. While these aspects are crucial to resemble detailed characteristics of experimental measurements, they are expected to not significantly influence the general accuracy and robustness of the numerical formulation, which is in the focus of the present study.} For reasons of numerical efficiency, the gas phase is approximated by an reduced density and dynamic viscosity ratio of $\rho_0^l/\rho_0^g=100$ and $\eta^l/\eta^g=10$ as analyzed in the last section. As thermal properties $c_p=10 J \, kg^{-1} \, K^{-1}$, $k=0.026 W\, m^{-1} \, K^{-1}$ and $\chi_l=0$ are chosen for the gas phase. Unless stated otherwise, the following standard discretization and regularization strategy is applied: The reference pressure of the weakly compressible model is set to $p_0=1.0 \cdot 10^7 N m^{-2}$ and the background pressure of the transport velocity formulation to $p_b=5p_0$ for both phases. Moreover, a standard time step size of $\Delta t = 10^{-6} ms$ and an initial particle spacing of $\Delta x = 5/3 \mu m$ is typically applied. Finally, the considered computational domains are surrounded by three layers of boundary particles with prescribed temperature $\hat{T}$ as well as free-slip boundary conditions for the fluid field.

\subsubsection{Laser heating of melt drop on solid substrate} \label{sec:numex_heateddrop}
\begin{figure}[htbp]
\centering
\subfigure [Static equilibrium configuration: surface tension gradient coefficient $\alpha'_0/ \rho_0=0.0$. Analytical solution in grey.]
{
\includegraphics[width=0.95\textwidth]{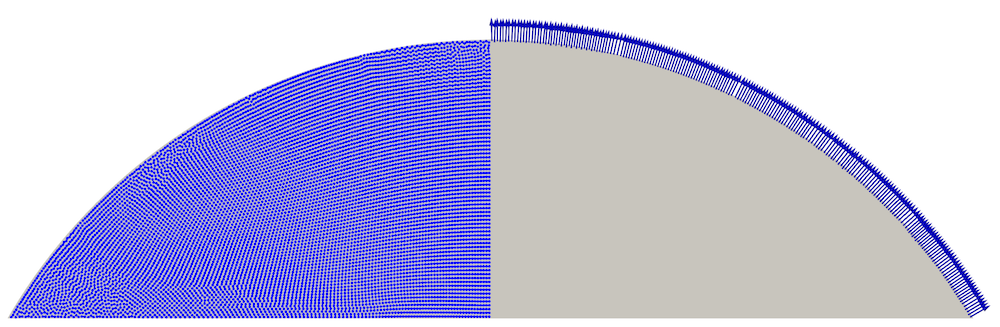}%
\label{fig:example3_1}
}
\subfigure [Low Marangoni convection strength: surface tension gradient coefficient $\alpha'_0/ \rho_0=2.0 \cdot 10^{-8}$.]
{
\includegraphics[width=0.95\textwidth]{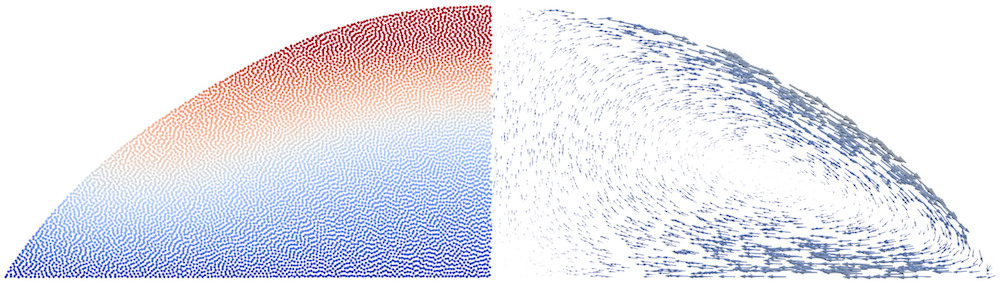}%
\label{fig:example3_2}
}
\subfigure [Medium Marangoni convection strength: surface tension gradient coefficient $\alpha'_0/ \rho_0=5.0 \cdot 10^{-8}$.]
{
\includegraphics[width=0.95\textwidth]{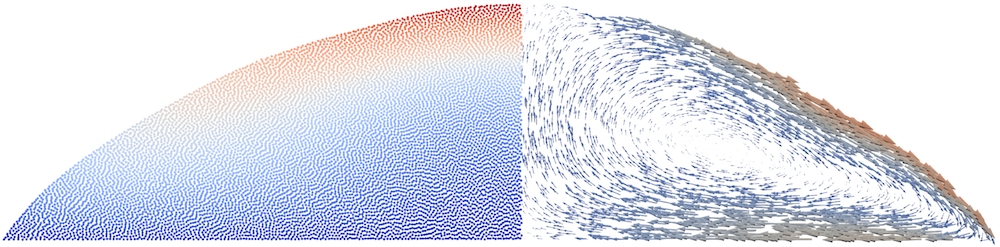}%
\label{fig:example3_3}
}
\subfigure [High Marangoni convection strength: surface tension gradient coefficient $\alpha'_0/ \rho_0=10.0 \cdot 10^{-8}$.]
{
\includegraphics[width=0.95\textwidth]{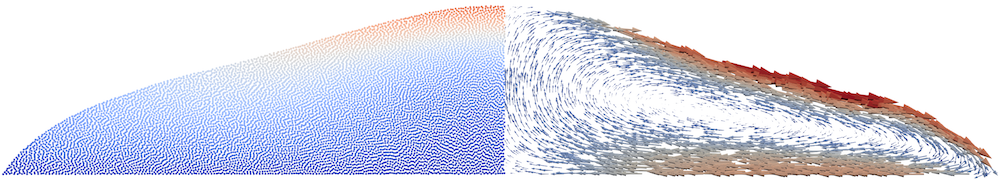}%
\label{fig:example3_4}
}
\caption{Laser heating of melt drop: Melt drop shape, temperature and velocity field for static contact angle $\theta_0=60^o$ and different Marangoni convection strengths ($\alpha'_0/ \rho_0$ in [$(N \, m^{-1} \, K^{-1})/(kg \, m^{-3})$]). Temperature range (left) from $1700K$ (blue) to $2800K$ (red). Velocity range (right) from $0.0 m/s$ (blue) to $4.2 m/s$ (red). Gas and boundary phase not displayed.}
\label{fig:example3}
\end{figure}
In this example, the effects already considered in Section~\ref{sec:numex_migratingbubble} shall be extended by wetting forces and a laser beam heat source. Thereto, a rectangular domain with dimensions $x \in [-100;100], \, y \in [-40;40]$ (all length dimensions given in $\mu m$) is considered. The sub-domain $x \in [-50;50], \, y \in [-40;-10]$ is initially covered with liquid melt, representing a droplet of initially rectangular shape, the remaining domain with gas. The overall domain is surrounded by three layers of boundary particles. In this example, the initial as well as the wall temperature (Dirichlet boundary conditions) have been chosen to $T_0=\hat{T}=1700K$. Note that this choice for the temperature initial and boundary conditions ensures that the melt drop remains liquid at all times and no phase change problem has to be considered in this example. The surface tension coefficient $\alpha_0$ is linearly increased from zero to $1.8 N \, m^{\!-1}$ in the time interval $t \in [0;0.05]$ (all time units in $ms$) and then kept constant throughout the rest of the simulation. Moreover, a static wetting angle of $\theta_0=60^o$ is considered. In the time interval $t \in [0;0.3]$, the laser heat source is set to zero such that the system can reach a static equilibrium configuration under the action of surface tension and wetting forces (see Figure~\ref{fig:example3_1}). In order to enable a fast transition into the equilibrium state, the dynamic viscosity of the melt phase is increased by a factor of ten compared to its standard value (only) in this first time interval $t \in [0;0.3]$. In the remaining simulation time $t \in [0.3;0.6]$, the surface tension, the dynamic viscosity as well as all the other material parameters are kept constant at their standard values according to Table~\ref{tab:material_params}. Additionally, the laser beam heat source is (instantaneously) switched on at $t=0.3$. The laser beam points into negative $y-$direction and its center is located at $x=0$. The irradiance is set to $s^{lg}_{l0} \approx 5.94 \cdot 10^{-3} W \, \mu m^{-2}$ ($s^{lg}_{l0}/\rho_0 = 8 \cdot 10^5 W \, m \, kg^{-1}$) and the characteristic radius to $r_w=40 \mu m$. The reference pressure of the weakly compressible model is set to $p_0= 1.0 \cdot 10^7 N m^{-2}$ for the melt and to $p_0=1.0 \cdot 10^6 N m^{-2}$ for the gas phase. The background pressure of the transport velocity formulation is set to $p_b=p_0$ for both phases. Moreover, a time step size of $\Delta t = 5.0 \cdot 10^{-7} ms$ and an initial particle spacing of $\Delta x = 5/12 \mu m$ is applied.\\
The analytical solution for the shape of the melt domain at the equilibrium state $t=0.3 ms$ is given by a circular segment as illustrated in Figure~\ref{fig:example3_1} in grey color. The numerical solution to this problem is characterized by a highly regular particle distribution and interface normal vector field, which represent the analytical solution with high accuracy (see also Figure~\ref{fig:example3_1}). Figures~\ref{fig:example3_2}-\ref{fig:example3_4} represent the steady state solutions resulting from the laser heating at $t \approx 0.6 ms$ for different choices of the surface tension gradient coefficient $\alpha'_0$ according to~\eqref{eq:fluid_surfacetension_tempdepend}. For simplicity, the ratio $\alpha'_0 / \rho_0$ has been prescribed in the simulations, where the four different variants $\alpha'_0 / \rho_0 = \{0.0, 2.0 \cdot 10^{-8}, 5.0 \cdot 10^{-8}, 10.0 \cdot 10^{-8}\}$ in $[(N \, m^{-1} \, K^{-1})/(kg \, m^{-3})]$ (corresponding to $\alpha'_0 = \{0.0, 1.486 \cdot 10^{-4}, 3.715 \cdot 10^{-4}, 7.43 \cdot 10^{-4}\}$ in $[N \, m^{-1} \, K^{-1}]$) have been analyzed. It can be observed that an increasing surface tension gradient coefficient (from Figure~\ref{fig:example3_2} to \ref{fig:example3_4}) leads to increased Marangoni convection characterized by higher velocity magnitudes and flatter droplet shapes. The combination of very flat droplet shapes and the static wetting angle constraint at the triple points even induces local regions with concave interface curvature for the highest surface tension gradient coefficient $\alpha'_0$ under consideration (Figure~\ref{fig:example3_4}). Besides the flatter droplet shape a higher Marangoni convection also results in a more homogeneous temperature distributions with lower peak temperature values. Both effects, i.e. flat droplets with homogeneous temperature distribution, can be considered as beneficial for typical additive manufacturing processes since problems related to non-smooth surface patterns (e.g. the balling effect due to surface tension-driven Plateau-Rayleigh instabilities~\cite{gusarov2010}) and overheating (e.g. excessive evaporation and keyhole instabilities~\cite{King2014}) are expected to be less pronounced.

\subsubsection{2D laser melting with wetting and Marangoni forces} \label{sec:numex_2Dmelt_wetmar}

\begin{figure}[htbp]
\centering
%
\subfigure [$\alpha'_0/ \rho_0\!=\!0.0$, $t\!=\!0.2ms$.]
{
\includegraphics[width=0.23\textwidth]{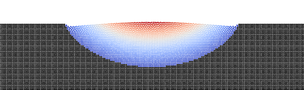}
}
\subfigure [$\alpha'_0/ \rho_0\!=\!0.0$, $t\!=\!0.5ms$.]
{
\includegraphics[width=0.23\textwidth]{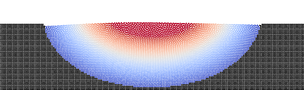}
}
\subfigure [$\alpha'_0/ \rho_0\!=\!0.0$, $t\!=\!0.6ms$.]
{
\includegraphics[width=0.23\textwidth]{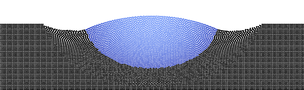}
}
\subfigure [$\alpha'_0/ \rho_0\!=\!0.0$, $t\!=\!0.7ms$.]
{
\includegraphics[width=0.23\textwidth]{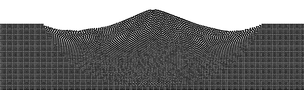}
}
\\
\subfigure [$\alpha'_0/ \rho_0\!=\!2.0$, $t\!=\!0.2ms$.]
{
\includegraphics[width=0.22\textwidth]{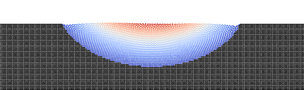}
}
\subfigure [$\alpha'_0/ \rho_0\!=\!2.0$, $t\!=\!0.5ms$.]
{
\includegraphics[width=0.23\textwidth]{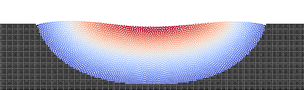}
}
\subfigure [$\alpha'_0/ \rho_0\!=\!2.0$, $t\!=\!0.6ms$.]
{
\includegraphics[width=0.23\textwidth]{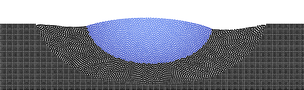}
}
\subfigure [$\alpha'_0/ \rho_0\!=\!2.0$, $t\!=\!0.7ms$.]
{
\includegraphics[width=0.23\textwidth]{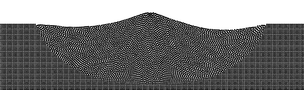}
}
\\
\subfigure [$\alpha'_0/ \rho_0\!=\!5.0$, $t\!=\!0.2ms$.]
{
\includegraphics[width=0.23\textwidth]{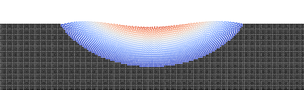}
}
\subfigure [$\alpha'_0/ \rho_0\!=\!5.0$, $t\!=\!0.5ms$.]
{
\includegraphics[width=0.23\textwidth]{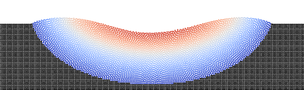}
}
\subfigure [$\alpha'_0/ \rho_0\!=\!5.0$, $t\!=\!0.6ms$.]
{
\includegraphics[width=0.23\textwidth]{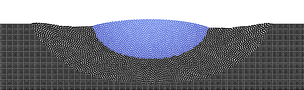}
}
\subfigure [$\alpha'_0/ \rho_0\!=\!5.0$, $t\!=\!0.7ms$.]
{
\includegraphics[width=0.23\textwidth]{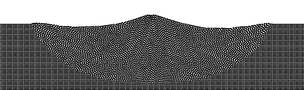}
}
\\
\subfigure [$\alpha'_0/ \rho_0\!=\!10.0$, $t\!=\!0.2ms$.]
{
\includegraphics[width=0.23\textwidth]{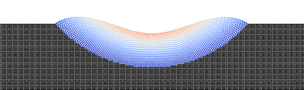}
}
\subfigure [$\alpha'_0/ \rho_0\!=\!10.0$, $t\!=\!0.5ms$.]
{
\includegraphics[width=0.23\textwidth]{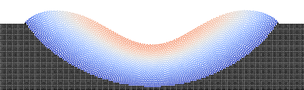}
}
\subfigure [$\alpha'_0/ \rho_0\!=\!10.0$, $t\!=\!0.6ms$.]
{
\includegraphics[width=0.23\textwidth]{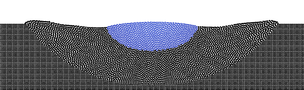}
}
\subfigure [$\alpha'_0/ \rho_0\!=\!10.0$, $t\!=\!0.7ms$.]
{
\includegraphics[width=0.23\textwidth]{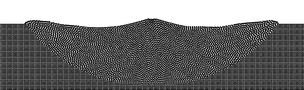}
}
%
\caption{2D laser melting with wetting and Marangoni forces: Different Marangoni convection strengths ($\alpha'_0/ \rho_0$ in $[10^{-8} (N \, m^{-1} \, K^{-1})/(kg \, m^{-3})]$) and time steps. Static wetting angle $\theta_0  = 75^o$. Temperature range from $1700K$ (blue) to $3500K$ (red). Solid phase in black. Gas and boundary particles not displayed.}
\label{fig:example4_0}
\end{figure}
\begin{figure}[b!!!]
\centering
%
\subfigure [$\theta_0  = 30^o$, $t=0.2ms$.]
{
\includegraphics[width=0.23\textwidth]{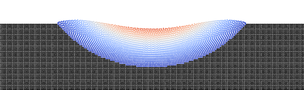}
}
\subfigure [$\theta_0  = 30^o$, $t=0.5ms$.]
{
\includegraphics[width=0.23\textwidth]{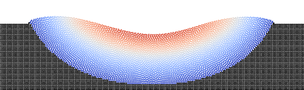}
}
\subfigure [$\theta_0  = 30^o$, $t=0.6ms$.]
{
\includegraphics[width=0.23\textwidth]{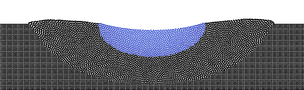}
}
\subfigure [$\theta_0  = 30^o$, $t=0.7ms$.]
{
\includegraphics[width=0.23\textwidth]{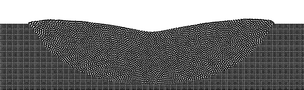}
}
\\
\subfigure [$\theta_0  = 60^o$, $t=0.2ms$.]
{
\includegraphics[width=0.23\textwidth]{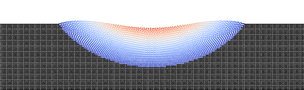}
}
\subfigure [$\theta_0  = 60^o$, $t=0.5ms$.]
{
\includegraphics[width=0.23\textwidth]{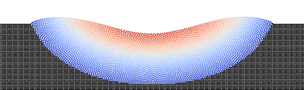}
}
\subfigure [$\theta_0  = 60^o$, $t=0.6ms$.]
{
\includegraphics[width=0.23\textwidth]{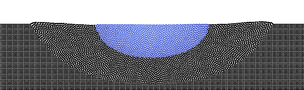}
}
\subfigure [$\theta_0  = 60^o$, $t=0.7ms$.]
{
\includegraphics[width=0.23\textwidth]{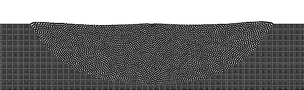}
}
\\
\subfigure [$\theta_0  = 75^o$, $t=0.2ms$.]
{
\includegraphics[width=0.23\textwidth]{figures/example4_wetangle75_marangoni05_t2_00}
}
\subfigure [$\theta_0  = 75^o$, $t=0.5ms$.]
{
\includegraphics[width=0.23\textwidth]{figures/example4_wetangle75_marangoni05_t5_00}
}
\subfigure [$\theta_0  = 75^o$, $t=0.6ms$.]
{
\includegraphics[width=0.23\textwidth]{figures/example4_wetangle75_marangoni05_t6_00}
}
\subfigure [$\theta_0  = 75^o$, $t=0.7ms$.]
{
\includegraphics[width=0.23\textwidth]{figures/example4_wetangle75_marangoni05_t7_00}
}
\\
\subfigure [$\theta_0  = 90^o$, $t=0.2ms$.]
{
\includegraphics[width=0.23\textwidth]{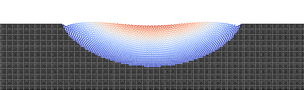}
}
\subfigure [$\theta_0  = 90^o$, $t=0.5ms$.]
{
\includegraphics[width=0.23\textwidth]{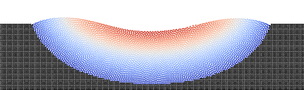}
}
\subfigure [$\theta_0  = 90^o$, $t=0.6ms$.]
{
\includegraphics[width=0.23\textwidth]{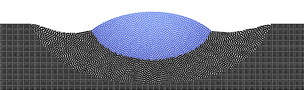}
}
\subfigure [$\theta_0  = 90^o$, $t=0.7ms$.]
{
\includegraphics[width=0.23\textwidth]{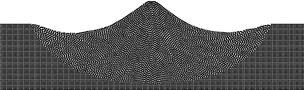}
}
%
\caption{2D laser melting with wetting and Marangoni forces: Different static wetting angles and time steps. Surface tension gradient coefficient $\alpha'_0/ \rho_0 = 5.0 \cdot 10^{-8} (N \, m^{-1} \, K^{-1})/(kg \, m^{-3})$. Temperature range (left) from $1700K$ (blue) to $3500K$ (red). Solid phase in black. Gas and boundary particles not displayed.}
\label{fig:example4_1}
\end{figure}
\begin{figure}[htbp]
\centering
\subfigure [Low Marangoni convection strength: surface tension gradient coefficient $\alpha'_0/ \rho_0=2.0$.]
{
\includegraphics[width=1.0\textwidth]{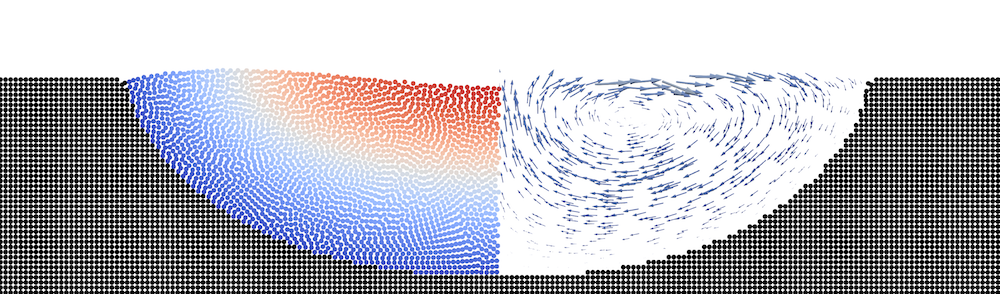}
\label{fig:example4_2a}
}
\\
\subfigure [Medium Marangoni convection strength: surface tension gradient coefficient $\alpha'_0/ \rho_0=5.0$.]
{
\includegraphics[width=1.0\textwidth]{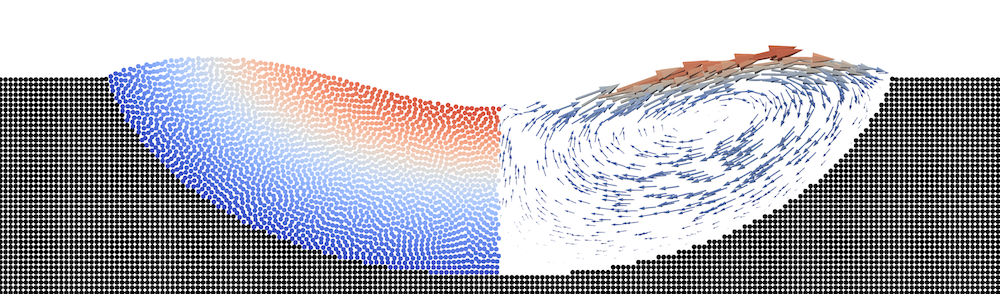}
\label{fig:example4_2b}
}
\\
\subfigure [High Marangoni convection strength: surface tension gradient coefficient $\alpha'_0/ \rho_0=10.0$.]
{
\includegraphics[width=1.0\textwidth]{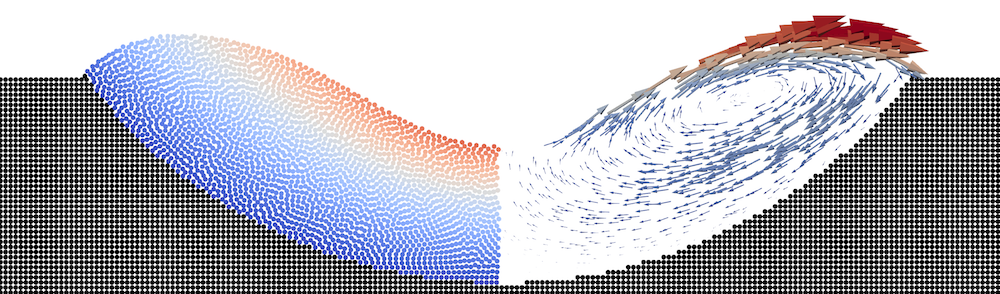}
\label{fig:example4_2c}
}
\caption{2D laser melting with wetting and Marangoni forces: Time step $t=0.5ms$. Static wetting angle $\theta_0  = 75^o$. Different Marangoni convection strengths ($\alpha'_0/ \rho_0$ in $[10^{-8} (N \, m^{-1} \, K^{-1})/(kg \, m^{-3})]$). Temperature range from $1700K$ (blue) to $3500K$ (right). Velocity range (right) from $0.0m/s$ to $0.8m/s$. Solid phase in black. Gas and boundary particles not displayed.}
\label{fig:example4_2}
\end{figure}
In a next step, the system considered in the last example is extended by the phase change problem. Thereto, the problem setup will be changed such that a solid domain exists initially instead of a melt phase domain. Specifically, a rectangular domain with dimensions $x \in [-100;100], \, y \in [-60;60]$ (all length dimensions given in $\mu m$) is considered. The lower half of the domain, i.e. $x \in [-100;100], \, y \in [-60; 0]$ is initially covered with a solid phase, the remaining domain with gas. The overall domain is surrounded by three layers of boundary particles. The initial as well as the wall temperature (Dirichlet boundary conditions) have been chosen to $T_0=\hat{T}=500K$. In the time interval $t \in [0;0.5]$ (all time units given in $ms$), a laser heat source ($s^{lg}_{l0} \approx 7.43 \cdot 10^{9} W \, m^{-2}$, i.e. $s^{lg}_{l0}/\rho_0 = 1.0 \cdot 10^6 W \, m \, kg^{-1}$; $r_w=60 \mu m$; located at $x_0=0$; pointing in negative $y-direction$) is acting onto the surface of the melt and solid material. In the time interval $t \in [0.5;1.0]$ the laser beam is switched off to study the subsequent solidification problem. Since melting is relevant in this test case, an additional viscous force contribution at the solid-liquid interface according to~\eqref{eq:sph_dissipation_sl} with the dimensionless viscosity parameter ${\zeta}^{sl}_0=10$ as well as $T_{max}=3500K$ applied. Note that for this specific example an untypically high value of $T_{max}$ has been chosen to increase the overall viscosity within a large portion of the melt pool, which in turn yields a laminar flow and a smooth velocity profile. This parameter setting has proven beneficial to analyze and isolate the influence of the static wetting angle and of the surface tension gradient coefficient on the resulting flow patterns. All remaining material parameters are chosen according to Table~\ref{tab:material_params}. The reference pressure of the weakly compressible model is set to $p_0=1.0 \cdot 10^7 N m^{-2}$ for the melt and to $p_0=1.0 \cdot 10^6 N m^{-2}$ for the gas phase. The background pressure of the transport velocity formulation is set to $p_b=p_0$ for both phases. Moreover, a time step size of $\Delta t = 2.5 \cdot 10^{-6} ms$ and an initial particle spacing of $\Delta x = 5/6 \mu m$ is applied. In Figure~\ref{fig:example4_0}, the melting and solidification process is displayed at different time instances and for different values of the surface tension gradient coefficient $\alpha'_0$ dictating the strength of the Marangoni convection. As already observed in the last example, the Marangoni convection fosters material transport from the melt pool center (high temperatures) to the edge of the melt pool (lower temperatures) leading to a depression at the melt pool center. Moreover, Marangoni convection (from first to fourth row of Figure~\ref{fig:example4_0}) is an effective means of heat transfer and reduces temperature gradients with increasing surface tension gradient coefficient $\alpha'_0$. During the solidification process (third and fourth column of Figure~\ref{fig:example4_0}) the triple point solid-liquid-gas moves with the solidification front towards the center of the melt pool. The wetting forces, tending to enforce the static wetting angle at the current triple point position, induce an upward warping of the liquid-gas interface while the triple point moves towards the melt pool center. As expected, the resulting peak at the center of the solidified melt pool is less pronounced for the higher values of $\alpha'_0$, since the initial depression of the liquid melt pool due to the Marangoni convection (partly) compensates this warping effect.\\
Similarly, in Figure~\ref{fig:example4_1}, the melting and solidification process is displayed at different time instances and for different values of the static wetting angle $\theta_0$. When comparing the different rows in the first and second column of Figure~\ref{fig:example4_1}, only minor differences can be observed in the resulting melt pool shapes. This observation can be explained by the fact that in all considered cases the triple point lies on (or is very close to) a sharp edge of the solid domain for which no unique normal vector $\vectorbold{n}^{sf}$ can be defined. On the other hand, during the solidification of the melt pool the same warping effect of the liquid-gas interface as in Figure~\ref{fig:example4_0} can be observed. As expected the dimension of the resulting peak at the center of the solidified melt pool increases with increasing values of the static wetting angle $\theta_0$.\\
Eventually, in Figure~\ref{fig:example4_2} also the resulting velocity profiles at $t=0.5ms$ are illustrated for the three different values of the surface tension gradient coefficient $\alpha'_0$ as considered above. The resulting velocity profiles are similar to the ones already observed in Section~\ref{sec:numex_heateddrop} with increasing velocity magnitudes for increasing values of the surface tension gradient coefficient $\alpha'_0$. {While the accuracy of the wetting as well as normal and tangential surface tension force contributions has already been verified individually in Sections~\ref{sec:numex_migratingbubble} and~\ref{sec:numex_droplet_oscillation} (and Figure~\ref{fig:example3_1} of the last section), the remaining examples in this and the last section mainly demonstrate the robustness of the numerical scheme when all these interface forces occur simultaneously. Specifically, the example of this section considers melting and solidification, i.e. a dynamic change of the size/shape of the liquid domain, as additional complexity. While the observed influence of wetting and Marangoni effects on the melt pool shape and dynamics is plausible (see discussion above), a comparison with experimentally observed behavior will be considered in our ongoing research work as additional means of validation.}

\subsubsection{2D laser melting with recoil pressure} \label{sec:numex_2Dmelt_rec}

\begin{figure}[b!!!]
\centering
\subfigure [Exemplary melt pool shape from laser melting experiment; taken from~\cite{cunningham2019keyhole}.]
{
\includegraphics[width=0.31\textwidth]{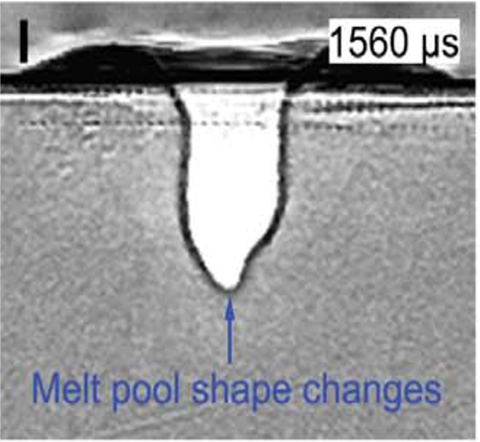}
\label{fig:example5_0a0}
}
\subfigure [Simulation with SPH model: particle distribution \textit{with} interface viscosity.]
{
\includegraphics[width=0.31\textwidth]{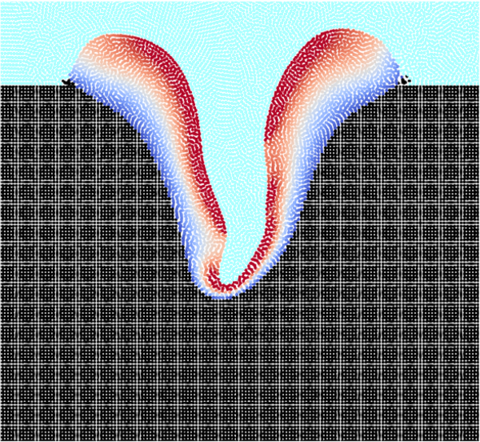}
\label{fig:example5_0a1}
}
\subfigure [Simulation with SPH model: particle distribution \textit{without} interface viscosity.]
{
\includegraphics[width=0.31\textwidth]{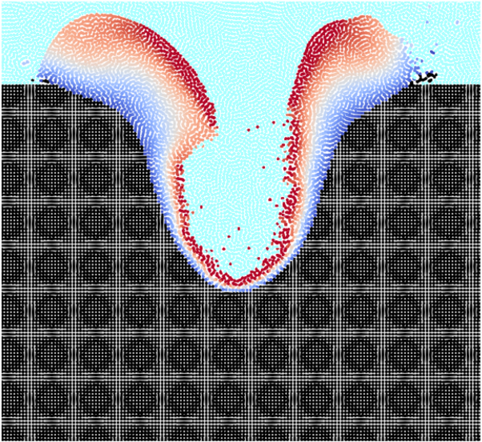}
\label{fig:example5_0a2}
}
\caption{{Melt pool shape from experiment (left). Simulation results for example "2D laser melting with recoil pressure" (\textit{variant 1}) at $t=0.072ms$: Visualization of spurious interface currents and stabilization via viscous interface forces (middle and right). Temperature range from $1700K$ (blue) to $3500K$ (red). Display of solid phase in black and gas phase in light blue.}}
\label{fig:example5_0a}
\end{figure}

The effects studied in Section~\ref{sec:numex_2Dmelt_wetmar} will now be extended by evaporation-induced recoil pressure forces according to~\eqref{eq:fluid_recoil}. In order to study deep keyhole dynamics in the range of high recoil pressure magnitudes, a larger domain with dimensions $x \in [-200;200], \, y \in [-200;200]$ (length dimensions given in $\mu m$) is considered. The lower region with $x \in [-200;200], \, y \in [-200; 40]$ is initially covered with a solid phase, the remaining domain with gas. The overall domain is surrounded by three layers of boundary particles. The initial as well as the wall temperature (Dirichlet boundary conditions) have been chosen to $T_0=\hat{T}=500K$. A laser heat source ($s^{lg}_{l0} \approx 16.0 W \, \mu m^{-2}$; $r_w=30 \mu m$; located at $x_0=0$; pointing in negative $y-$direction) is acting onto the surface of the melt and solid material. Again, an additional viscous force contribution at the solid-liquid interface according to~\eqref{eq:sph_dissipation_sl} with the dimensionless viscosity parameter ${\zeta}^{sl}_0=5$ as well as $T_{max}=2000K$ is applied. Moreover, to reduce numerical instabilities at the liquid-gas interface due to the high recoil pressure forces acting in this example, at this interface an additional viscous force according to~\eqref{eq:sph_dissipation_lg} with ${\zeta}^{lg}_0=2.5 \cdot 10^{-4}$ is applied. To isolate the effect of surface tension-recoil pressure interaction, no wetting and Marangoni forces are considered. All remaining material parameters are chosen according to Table~\ref{tab:material_params}. The reference pressure of the weakly compressible model is set to $p_0=1.0 \cdot 10^7 N m^{-2}$ for the melt and gas phase. 
\begin{figure}[b!!!]
\centering
\begin{minipage}{0.95\textwidth}
\centering
\subfigure [Particle distribution \textit{with} interface viscosity.]
{
\includegraphics[width=0.45\textwidth]{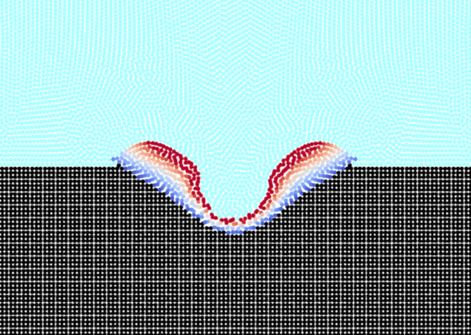}
\label{fig:example5_0b1}
}
\subfigure [Particle distribution \textit{without} interface viscosity.]
{
\includegraphics[width=0.45\textwidth]{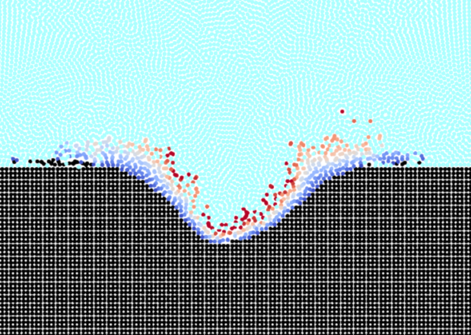}
\label{fig:example5_0b2}
}
\subfigure [Velocity field \textit{with} interface viscosity.]
{
\includegraphics[width=0.45\textwidth]{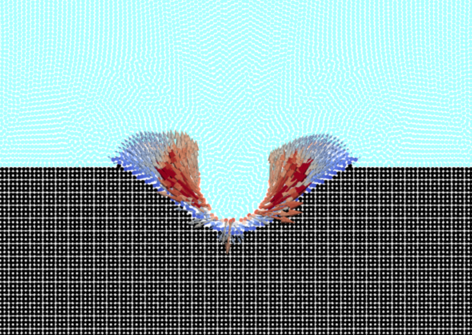}
\label{fig:example5_0b3}
}
\subfigure [Velocity field \textit{without} interface viscosity.]
{
\includegraphics[width=0.45\textwidth]{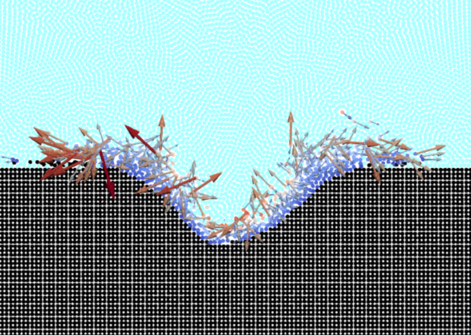}
\label{fig:example5_0b4}
}
\caption{2D laser melting with recoil pressure according to \textit{variant 2} at time $t=0.0255ms$: Visualization of spurious interface currents and stabilization via viscous interface forces. Temperature range from $1700K$ (blue) to $3500K$ (red). Velocity range from $0ms^{-1}$ (blue) to $10ms^{-1}$ (red). Display of solid phase in black and gas phase in light blue. Note that only a subdomain with $x \in [-100;100], \, y \in [-100; 100]$ is displayed.}
\label{fig:example5_0b}
\end{minipage}
\end{figure}
The background pressure of the transport velocity formulation is set to $p_b=5p_0$ for both phases. Moreover, a time step size of $\Delta t = 1.0 \cdot 10^{-6} ms$ and an initial particle spacing of $\Delta x = 5/3 \mu m$ is applied. The problem setup described by these parameters is denoted as \textit{variant 1} in the following. In order to study the influence of the evaporative mass loss~\eqref{eq:fluid_evaporation} a second variant (\textit{variant 2}) of this problem is considered where this term is neglected (while the recoil pressure is still considered). In order to end up with comparable thermal characteristics the laser irradiance has been decreased to a value of $s^{lg}_{l0} \approx 0.08 W \, \mu m^{-2}$ for \textit{variant 2}, which turned out to result in comparable peak temperatures in the melt pool center. Due to the higher interface dynamics in \textit{variant 2} the viscosity factor in~\eqref{eq:sph_dissipation_lg} is increased to ${\zeta}^{lg}_0=1.0 \cdot 10^{-3}$. Moreover, for \textit{variant 2} a smaller domain with dimensions $x \in [-200;200], \, y \in [-150; 150]$ is considered.\\
To study the influence of the viscous interface force, simulations with and without this additional stabilization term shall be compared. In Figure~\ref{fig:example5_0a} the resulting melt pool shape at $t=0.072ms$ is displayed for these two cases and problem \textit{variant 1}. Without interface stabilization term (Figure~\ref{fig:example5_0a2}) the spurious interface flows lead to a non-smooth, strongly distorted interface topology fostering nonphysical interface dynamics and eventually instability of the numerical scheme. In contrast, the additional interface stabilization term (Figure~\ref{fig:example5_0a1}) results in a smooth and physically reasonable interface topology by damping out the spurious interface flows and thus effectively stabilizing the numerical scheme. The high recoil pressure forces form a deep depression at the center of the melt pool, typically denoted as keyhole. Note that the configuration displayed in Figure~\ref{fig:example5_0a1} is no stationary configuration of the melt pool but rather characterized by high-frequency fluctuations of the liquid-gas interface especially at the bottom of the keyhole. {Qualitatively, Figure~\ref{fig:example5_0a1} shows good agreement with representative experimental results (see Figure~\ref{fig:example5_0a0}, taken from~\cite{cunningham2019keyhole}; note that the specific process parameters from experiment and simulation don't match.)}\\
In Figure~\ref{fig:example5_0b} the melt pool shape resulting from simulations with and without additional stabilization term is displayed for problem \textit{variant 2} at $t=0.0255ms$. The neglect of evaporative heat losses in problem \textit{variant 2} leads to stronger recoil pressure dynamics and, consequently, the spurious interface flows are even more pronounced. The result is a strongly distorted interface topology and velocity field as illustrated in Figures~\ref{fig:example5_0b2} and~\ref{fig:example5_0b4}. Again, the additional interface stabilization term results in a smooth and physically reasonable interface topology and velocity field as shown in Figures~\ref{fig:example5_0b1} and~\ref{fig:example5_0b3}.
\begin{figure}[htbp]
\centering
%
\subfigure [time: $t=0.079ms$]
{
\includegraphics[width=0.23\textwidth]{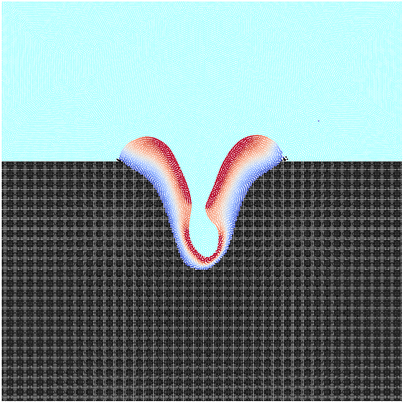}
\label{fig:example5_2_1}
}
\subfigure [time: $t=0.083ms$]
{
\includegraphics[width=0.23\textwidth]{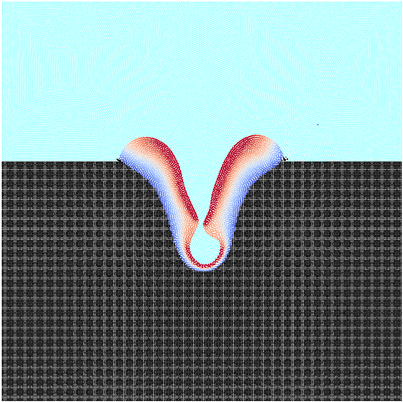}
\label{fig:example5_2_2}
}
\subfigure [time: $t=0.085ms$]
{
\includegraphics[width=0.23\textwidth]{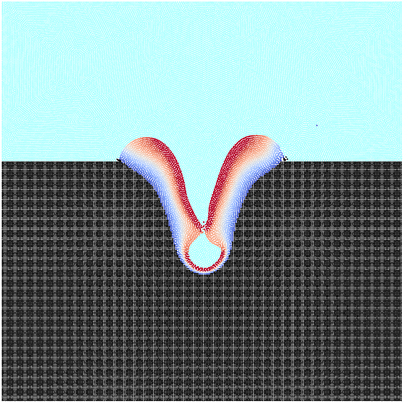}
\label{fig:example5_2_3}
}
\subfigure [time: $t=0.089ms$]
{
\includegraphics[width=0.23\textwidth]{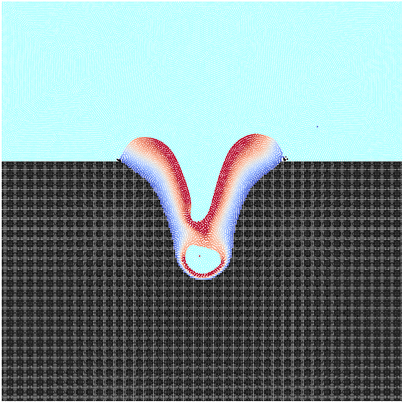}
\label{fig:example5_2_4}
}
\caption{2D laser melting with recoil pressure according to \textit{variant 1}: creation mechanism of a gas inclusion at different time steps. Temperature range from $1700K$ (blue) to $3500K$ (red). Display of solid phase in black and gas phase in light blue.}
\label{fig:example5_2}
\end{figure}
\begin{figure}[htbp]
\centering
%
\subfigure [time: $t=0.056ms$]
{
\includegraphics[width=0.23\textwidth]{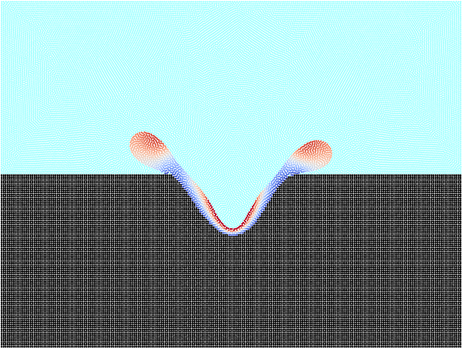}
\label{fig:example5_1_1}
}
\subfigure [time: $t=0.082ms$]
{
\includegraphics[width=0.23\textwidth]{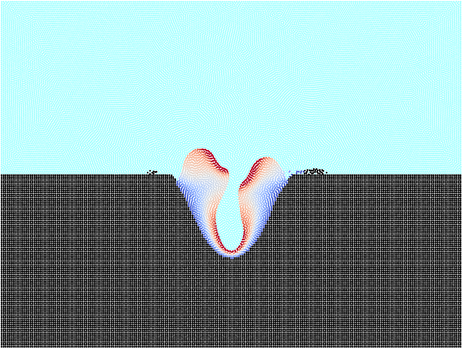}
\label{fig:example5_1_2}
}
\subfigure [time: $t=0.109ms$]
{
\includegraphics[width=0.23\textwidth]{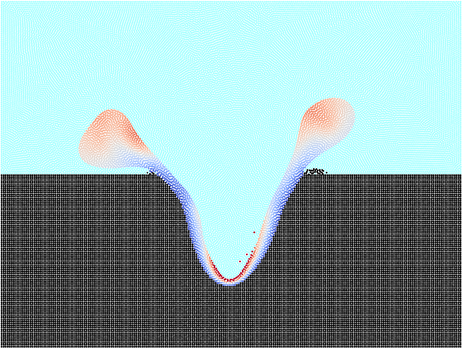}
\label{fig:example5_1_3}
}
\subfigure [time: $t=0.135ms$]
{
\includegraphics[width=0.23\textwidth]{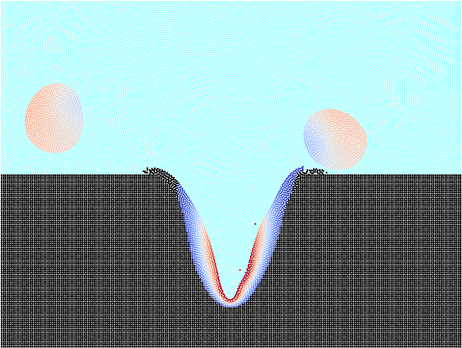}
\label{fig:example5_1_4}
}
\caption{2D laser melting with recoil pressure according to \textit{variant 2}: process of melt drop ejection at different time steps. Temperature range from $1700K$ (blue) to $3500K$ (red). Display of solid phase in black and gas phase in light blue.}
\label{fig:example5_1}
\end{figure}
Eventually, the melt pool thermo-hydrodynamics resulting from the two considered problem variants shall be studied for longer physical time spans. Figure~\ref{fig:example5_2} displays the system behavior of \textit{variant 1}. Accordingly, the interaction of surface tension and recoil pressure forces leads to oscillations of the liquid-gas interface with maximal amplitudes slightly above the bottom of the keyhole. Once, the amplitudes of these oscillations are large enough such that the opposite keyhole walls gets into touch a liquid bridge forms, which effectively encloses the gas material at the keyhole base - a gas inclusion is formed. Note that this example mainly aims at demonstrating the robustness of the proposed formulation in representing the highly dynamic surface tension-recoil pressure interaction and eventually the formation of a gas bubble. Of course, the employed phenomenological recoil pressure model does not explicitly resolve the high-velocity vapor jet that would arise from the keyhole base in the real physical system. The latter might considerably influence the described creation mechanism of a liquid bridge eventually resulting in a gas inclusion. Moreover, the present formulation does not apply any form of ray tracing for the laser heat source. Specifically, once the gas bubble has formed, the present model applies a heat source term not only on the upper interface of the arising liquid bridge but also at the bottom interface of the gas inclusion (which is visible in Figure~\ref{fig:example5_2_4} through the high peak temperatures at these locations). While an elimination of this deficiency (e.g. via ray tracing) can be considered as rather straight-forward process, the present formulation is still deemed suitable to demonstrate the robustness and general working principle of the proposed model and to study the problem-specific physical phenomena except for the subsequent motion of the closed gas bubble. Figure~\ref{fig:example5_1} displays the melt pool dynamics resulting from problem \textit{variant 2}. The high recoil pressure magnitudes in this variant result in periodic flow patterns consisting of recoil pressure-driven high-velocity waves from the center to the edges of the melt pool and surface tension-driven back flow from the edges to the center. The continued energy input by the laser beam heat sources results in increasing amplitudes of these flow oscillations until eventually melt droplets are ejected at the melt pool edges (Figures~\ref{fig:example5_2_3} and~\ref{fig:example5_2_4}). This example demonstrates that also the highly dynamic evolution of the interface topology during melt drop ejection can be captured by the proposed formulation in a robust manner.

\subsubsection{3D laser melting problem with explicitly resolved powder particle geometry} \label{sec:numex_3Dmelt}

\begin{figure}[htbp]
\centering
\subfigure [time: $0.0ms$]
{
\includegraphics[width=0.44\textwidth]{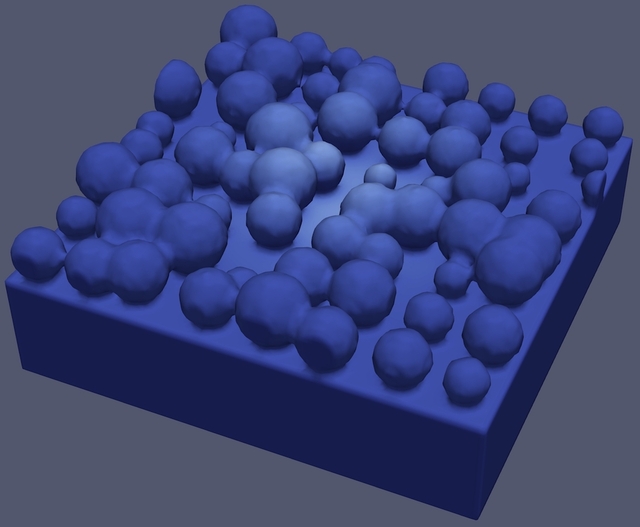}
\label{fig:example6_1a}
}
\subfigure [time: $0.080ms$]
{
\includegraphics[width=0.44\textwidth]{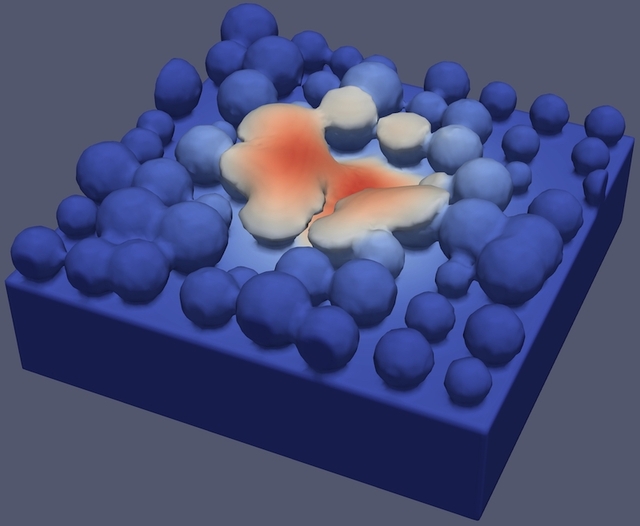}
\label{fig:example6_1b}
}
\subfigure [time: $0.180ms$]
{
\includegraphics[width=0.44\textwidth]{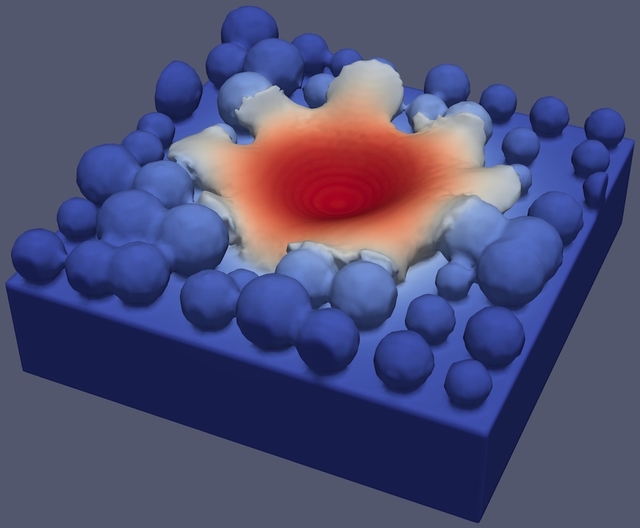}
\label{fig:example6_1c}
}
\subfigure [time: $0.455ms$]
{
\includegraphics[width=0.44\textwidth]{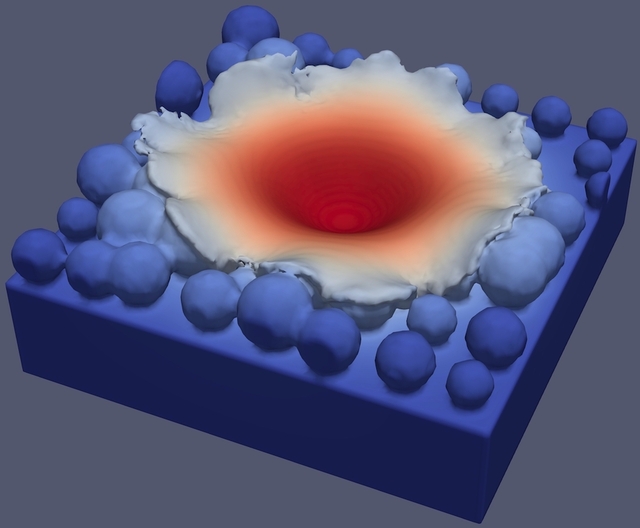}
\label{fig:example6_1d}
}
\subfigure [time: $0.630ms$]
{
\includegraphics[width=0.44\textwidth]{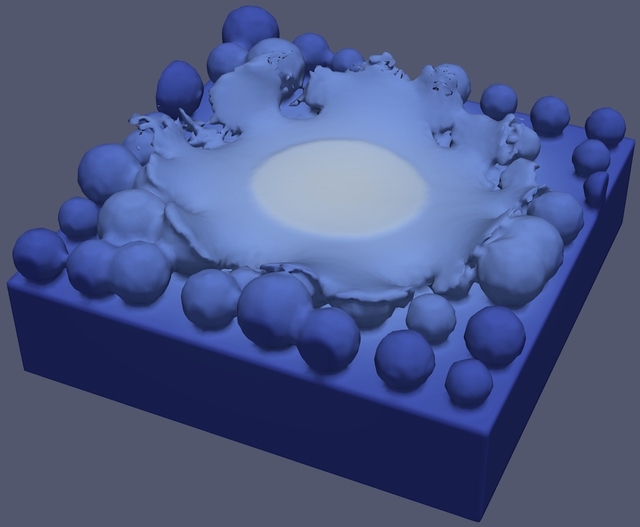}
\label{fig:example6_1e}
}
\subfigure [time: $1.000ms$]
{
\includegraphics[width=0.44\textwidth]{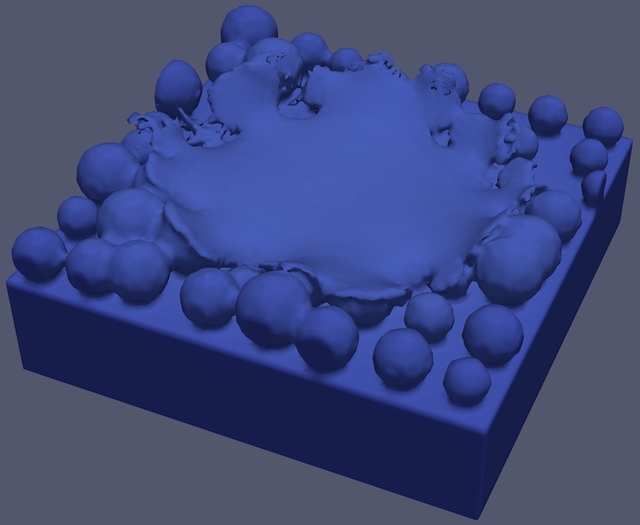}
\label{fig:example6_1f}
}
\caption{3D point melting problem with explicitly resolved powder particle geometry: resulting melt pool shape and final topology of solidified surface (post-processed) as well as temperature field in the range from $300K$ (blue) to $3700K$ (red).}
\label{fig:example6_1}
\end{figure}

In this last example, the problem considered in the last section shall be extended to 3D as well as more complex and realistic geometries by resolving individual particles of the metal powder in PBFAM as illustrated in Figure~\ref{fig:example6_1a}. The problem domain is confined by $x \in [-100;100], \, y \in [-100;100 ], \, z \in [-50;50]$ (length dimensions given in $\mu m$). The lower half with $x \in [-100;100], \, y \in [-100;100], \, z \in [-50;0],$ is initially covered with a solid phase, the remaining domain with gas. In addition, (spatially fixed) powder particles with diameters between $160 \mu m$ and $320 \mu m$ are randomly placed on top of the solid substrate. The overall domain is surrounded by three layers of boundary particles. The initial as well as the wall temperature (Dirichlet boundary conditions) have been chosen to $T_0=\hat{T}=500K$. A laser heat source ($s^{lg}_{l0} \approx 1.5 W \, \mu m^{-2}$; $r_w=60 \mu m$; located at $x_0=0$; pointing in negative $z-$direction) is acting onto the surface of the melt and solid material. The material parameters are chosen according to Table~\ref{tab:material_params}. The reference pressure of the weakly compressible model is set to $p_0=1.0 \cdot 10^7 N m^{-2}$ for the melt and gas phase. The background pressure of the transport velocity formulation is set to $p_b=5p_0$ for both phases. Moreover, a time step size of $\Delta t = 1.0 \cdot 10^{-6} ms$ and an initial particle spacing of $\Delta x = 5/3 \mu m$ is applied. The laser beam is switched on only for the time interval $t \in  [0;0.5]$ and the cooling / solidification process is considered in the remaining time interval $t \in  [0.5;1]$. Thus, in total the discrete problem consists of one million time steps and approximately one million SPH particles. Figure~\ref{fig:example6_1} illustrates the melting and solidification process at different time steps. Surface tension forces dominate volumetric forces such as gravity at the considered length scales and smoothen out the original particle contours almost immediately after melting (see Figures~\ref{fig:example6_1b} and~\ref{fig:example6_1c}). The peak temperatures at the melt pool center exceed the boiling temperature of the liquid metal with ongoing exposure time. The resulting recoil pressure forces foster an increasingly deep depression at the melt pool center as illustrated in Figures~\ref{fig:example6_1c} and~\ref{fig:example6_1d}. For the chosen set of parameters the time scales governing the surface tension forces dominate the time scales of (conductive) heat transfer. Thus, once the laser is switched off, the surface tension forces close the melt pool depression before the liquid metal solidifies again resulting in a smooth and rather plain surface topology at the former melt pool center (see Figure~\ref{fig:example6_1f}).

\begin{figure}[htbp]
\centering
\subfigure [time: approx.  $0.12ms$ \hspace{0.22\textwidth}]
{
\includegraphics[width=0.48\textwidth]{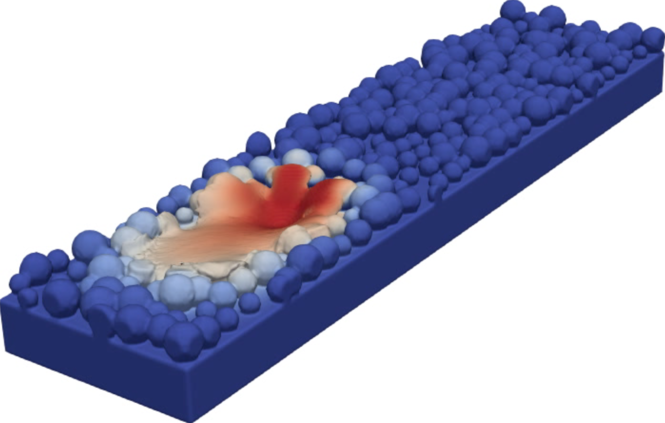}
\label{fig:example6_2a}
}
\subfigure [time: approx.  $0.24ms$ \hspace{0.22\textwidth}]
{
\includegraphics[width=0.48\textwidth]{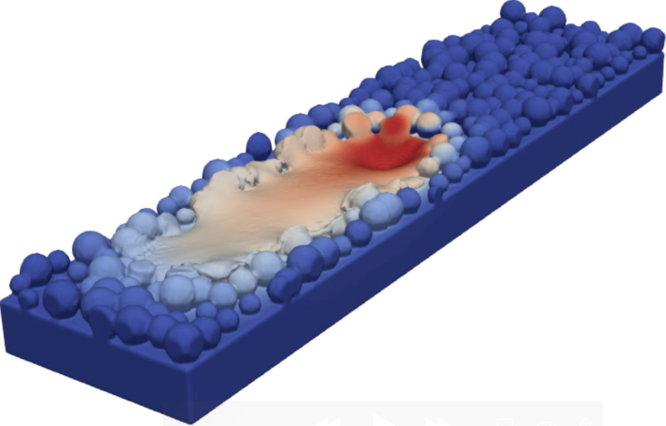}
\label{fig:example6_2b}
}
\subfigure [time: approx.  $0.36ms$ \hspace{0.22\textwidth}]
{
\includegraphics[width=0.48\textwidth]{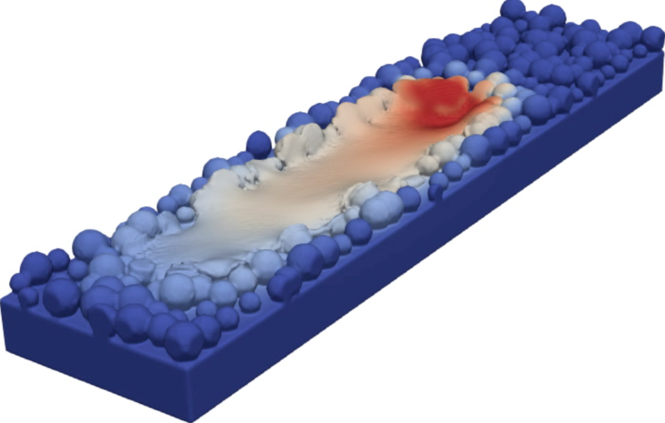}
\label{fig:example6_2c}
}
\subfigure [time: approx.  $0.48ms$ \hspace{0.22\textwidth}]
{
\includegraphics[width=0.48\textwidth]{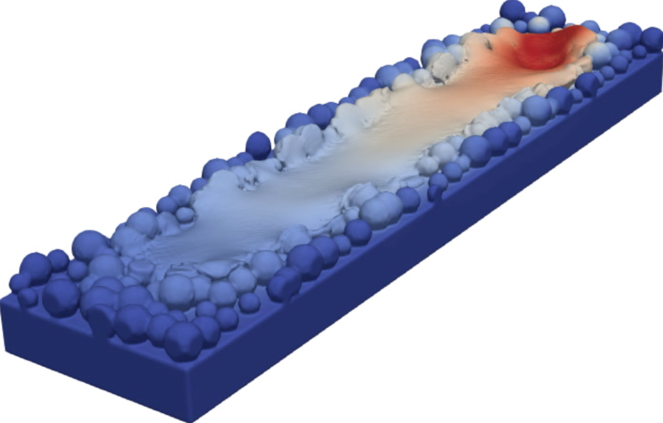}
\label{fig:example6_2d}
}
\caption{3D line melting problem with explicitly resolved powder particle geometry: resulting melt pool shape and final topology of solidified surface (post-processed) as well as temperature field in the range from $300K$ (blue) to $3700K$ (red).}
\label{fig:example6_2}
\end{figure}

Eventually, the example is extended to line scanning by increasing the total length of the domain by a factor of four in $x-$direction, increasing the laser energy density by a factor of two and moving the laser with a constant velocity of $v_x=1m/s$ in $x-$direction. The results are displayed in Figure~\ref{fig:example6_2}. All in all, both variants of this example, i.e. point and line scanning, demonstrate the robustness and general suitability of the proposed SPH formulation for 3D thermo-capillary phase change problems with geometrically complex features as occurring in PBFAM.

\section{Conclusion and Outlook} \label{sec:concl}

In the present work, a weakly compressible SPH formulation for thermo-capillary phase change problems involving solid, liquid and gaseous phases has been proposed with special focus on laser melting processes such as metal additive manufacturing. Evaporation-induced recoil pressure, temperature-dependent surface tension and wetting forces have been considered as mechanical interface fluxes, while a Gaussian laser beam heat source and evaporation-induced heat losses have been considered as thermal interface fluxes.\\
Specifically, a novel interface stabilization scheme based on viscous interface forces has been proposed, which was shown to effectively damp spurious interface flows well-known for continuum surface force approaches. It was demonstrated that only by this means recoil pressure-dominated problems, which typically arise in metal AM processes in the range of high laser powers, can be represented in a stable and physically accurate manner. Moreover, different SPH discretizations for the tangential projection of the temperature gradient, as required for the discrete Marangoni forces, have been reviewed. It was demonstrated analytically and numerically that standard \textit{two-sided} gradient approximations are sufficient for this purpose as long as zero-order consistency is satisfied, e.g. by anti-symmetric gradient construction.\\
In the context of metal AM melt pool modeling, the proposed formulation has been employed to study the influence of wetting forces and Marangoni convection of different strength on the resulting melt pool dynamics and shape as well as on the topology of the final solidified metal surface. Moreover, the robustness of the proposed scheme has been demonstrated by representing highly dynamic recoil-pressure-induced fluctuations and significant topology changes of the liquid-gas interface, e.g. during the ejection of liquid droplets from the melt pool or the inclusion of gas bubbles into the melt pool. Finally, the robustness, efficiency and general suitability of the proposed SPH formulation for 3D thermo-capillary phase change problems with geometrically complex features has been indicated by means of a 3D laser melting problem with explicit resolution of individual metal powder particles.

\section*{Acknowledgements} \label{sec:Acknowledgements}

This work was supported by a postdoc fellowship of the German Academic Exchange Service (DAAD) and by funding of the Deutsche Forschungsgemeinschaft (DFG, German Research Foundation) within project 437616465 and project 414180263. The authors would like to thank Lennart Schulze for his contributions in the code implementation and comparison of different temperature gradient discretization strategies and Yushen Sun for his support in the visualization of 3D melt pool simulations.

\bibliographystyle{elsarticle-num} 
\bibliography{collection.bib}

\begin{thebibliography}{10}
\expandafter\ifx\csname url\endcsname\relax
  \def\url#1{\texttt{#1}}\fi
\expandafter\ifx\csname urlprefix\endcsname\relax\def\urlprefix{URL }\fi
\expandafter\ifx\csname href\endcsname\relax
  \def\href#1#2{#2} \def\path#1{#1}\fi

\bibitem{Chang2015}
B.~Chang, C.~Allen, J.~Blackburn, P.~Hilton, D.~Du, {Fluid Flow Characteristics
  and Porosity Behavior in Full Penetration Laser Welding of a Titanium Alloy},
  Metallurgical and Materials Transactions B: Process Metallurgy and Materials
  Processing Science 46~(2) (2015) 906--918.

\bibitem{Geiger2009}
M.~Geiger, K.~H. Leitz, H.~Koch, A.~Otto, {A 3D transient model of keyhole and
  melt pool dynamics in laser beam welding applied to the joining of zinc
  coated sheets}, Production Engineering 3~(2) (2009) 127--136.

\bibitem{Ki2002a}
H.~Ki, J.~Mazumder, P.~S. Mohanty, {Modeling of laser keyhole welding: Part I.
  mathematical modeling, numerical methodology, role of recoil pressure,
  multiple reflections, and free surface evolution}, Metallurgical and
  Materials Transactions A 33~(June) (2002) 1817--1830.

\bibitem{Ki2002b}
H.~Ki, J.~Mazumder, P.~S. Mohanty, {Modeling of laser keyhole welding: Part II.
  simulation of keyhole evolution, velocity, temperature profile, and
  experimental verification}, Metallurgical and Materials Transactions A 33~(6)
  (2002) 1831--1842.

\bibitem{Rai2008}
R.~Rai, P.~Burgardt, J.~O. Milewski, T.~J. Lienert, T.~DebRoy, {Heat transfer
  and fluid flow during electron beam welding of 21Cr-6Ni-9Mn steel and
  Ti-6Al-4V alloy}, Journal of Physics D: Applied Physics 42~(2) (2008) 25503.

\bibitem{Semak1999}
V.~V. Semak, W.~D. Bragg, B.~Damkroger, S.~Kempka, {Transient model for the
  keyhole during laser welding}, Journal of Physics D: Applied Physics 32
  (1999) L61--L64.

\bibitem{Nugent2000}
S.~Nugent, H.~A. Posch, {Liquid drops and surface tension with smoothed
  particle applied mechanics}, Physical Review E - Statistical Physics,
  Plasmas, Fluids, and Related Interdisciplinary Topics 62~(4) (2000)
  4968--4975.

\bibitem{Tartakovsky2005}
A.~Tartakovsky, P.~Meakin, {Modeling of surface tension and contact angles with
  smoothed particle hydrodynamics}, Physical Review E - Statistical, Nonlinear,
  and Soft Matter Physics 72~(2) (2005) 1--9.

\bibitem{Tartakovsky2016}
A.~M. Tartakovsky, A.~Panchenko, {Pairwise Force Smoothed Particle
  Hydrodynamics model for multiphase flow: Surface tension and contact line
  dynamics}, Journal of Computational Physics 305 (2016) 1119--1146.

\bibitem{Brackbill1996}
J.~Brackbill, D.~Kothe, {Dynamic modeling of the surface tension}, in:
  Proceedings of the 3rd Microgravity Fluid Physics Conference, Cleveland, OH,
  1996, pp. 693--698.

\bibitem{Lafaurie1994}
B.~Lafaurie, C.~Nardone, R.~Scardovelli, S.~Zaleski, G.~Zanetti, {Modelling
  merging and fragmentation in multiphase flows with SURFER}, Journal of
  Computational Physics 113~(1) (1994) 134--147.

\bibitem{Hu2006}
X.~Y. Hu, N.~A. Adams, {A multi-phase SPH method for macroscopic and mesoscopic
  flows}, Journal of Computational Physics 213~(2) (2006) 844--861.

\bibitem{Morris2000}
J.~P. Morris, {Simulating surface tension with smoothed particle
  hydrodynamics}, International Journal for Numerical Methods in Fluids 33~(3)
  (2000) 333--353.

\bibitem{Adami2010}
S.~Adami, X.~Y. Hu, N.~A. Adams, {A new surface-tension formulation for
  multi-phase SPH using a reproducing divergence approximation}, Journal of
  Computational Physics 229~(13) (2010) 5011--5021.

\bibitem{Andersson2010}
B.~Andersson, S.~Jakobsson, A.~Mark, F.~Edelvik, L.~Davidson, Modeling surface
  tension in sph by interface reconstruction using radial basis functions, in:
  Proc. of the 5th International SPHERIC Workshop, Vol.~3, 2010.

\bibitem{Zhang2010}
M.~Zhang, {Simulation of surface tension in 2D and 3D with smoothed particle
  hydrodynamics method}, Journal of Computational Physics 229~(19) (2010)
  7238--7259.

\bibitem{Zhang2015a}
A.~Zhang, P.~Sun, F.~Ming, {An SPH modeling of bubble rising and coalescing in
  three dimensions}, Computer Methods in Applied Mechanics and Engineering
  294~(145) (2015) 189--209.

\bibitem{Zhang2015b}
M.~Zhang, X.~L. Deng, {A sharp interface method for SPH}, Journal of
  Computational Physics 302 (2015) 469--484.

\bibitem{Breinlinger2013}
T.~Breinlinger, P.~Polfer, A.~Hashibon, T.~Kraft, {Surface tension and wetting
  effects with smoothed particle hydrodynamics}, Journal of Computational
  Physics 243 (2013) 14--27.

\bibitem{Das2010}
A.~Das, P.~Das, {Equilibrium shape and contact angle of sessile drops of
  different volumes—Computation by SPH and its further improvement by DI},
  Chemical Engineering Science 65~(13) (2010) 4027--4037.

\bibitem{Tong2014}
M.~Tong, D.~J. Browne, {An incompressible multi-phase smoothed particle
  hydrodynamics (SPH) method for modelling thermocapillary flow}, International
  Journal of Heat and Mass Transfer 73 (2014) 284--292.

\bibitem{Hopp-Hirschler2018}
M.~Hopp-Hirschler, M.~S. Shadloo, U.~Nieken, {A Smoothed Particle Hydrodynamics
  approach for thermo-capillary flows}, Computers and Fluids 176 (2018) 1--19.

\bibitem{Russell2018}
M.~A. Russell, A.~Souto-Iglesias, T.~I. Zohdi, {Numerical simulation of Laser
  Fusion Additive Manufacturing processes using the SPH method}, Computer
  Methods in Applied Mechanics and Engineering 341 (2018) 163--187.

\bibitem{Wessels2018}
H.~Wessels, C.~Wei{\ss}enfels, P.~Wriggers, {Metal particle fusion analysis for
  additive manufacturing using the stabilized optimal transportation meshfree
  method}, Computer Methods in Applied Mechanics and Engineering 339 (2018)
  91--114.

\bibitem{trautmann2018numerical}
M.~Trautmann, M.~Hertel, U.~F{\"u}ssel, Numerical simulation of weld pool
  dynamics using a sph approach, Welding in the World 62~(5) (2018) 1013--1020.

\bibitem{Weirather2019}
J.~Weirather, V.~Rozov, M.~Wille, P.~Schuler, C.~Seidel, N.~A. Adams, M.~F.
  Zaeh, {A Smoothed Particle Hydrodynamics Model for Laser Beam Melting of
  Ni-based Alloy 718}, Computers and Mathematics with Applications 78~(7)
  (2019) 2377--2394.

\bibitem{shah2020simulations}
D.~Shah, A.~N. Volkov, Simulations of deep drilling of metals by continuous
  wave lasers using combined smoothed particle hydrodynamics and ray-tracing
  methods, Applied Physics A 126~(2) (2020) 1--12.

\bibitem{furstenau2020generating}
J.-P. F{\"u}rstenau, H.~Wessels, C.~Wei{\ss}enfels, P.~Wriggers, Generating
  virtual process maps of slm using powder-scale sph simulations, Computational
  Particle Mechanics 7~(4) (2020) 655--677.

\bibitem{dao2021simulations}
M.~H. Dao, J.~Lou, Simulations of laser assisted additive manufacturing by
  smoothed particle hydrodynamics, Computer Methods in Applied Mechanics and
  Engineering 373 (2021) 113491.

\bibitem{Meier2017}
C.~Meier, R.~W. Penny, Y.~Zou, J.~S. Gibbs, A.~J. Hart, {Thermophysical
  Phenomena in Metal Additive Manufacturing by Selective Laser Melting:
  Fundamentals, Modeling, Simulation and Experimentation}, Annual Review of
  Heat Transfer 20 (2017).

\bibitem{Khairallah2014}
S.~A. Khairallah, A.~Anderson, {Mesoscopic simulation model of selective laser
  melting of stainless steel powder}, Journal of Materials Processing
  Technology 214~(11) (2014) 2627--2636.

\bibitem{Khairallah2016}
S.~A. Khairallah, A.~T. Anderson, A.~Rubenchik, W.~E. King, {Laser powder-bed
  fusion additive manufacturing: Physics of complex melt flow and formation
  mechanisms of pores, spatter, and denudation zones}, Acta Materialia 108
  (2016) 36--45.

\bibitem{Khairallah2020}
S.~A. Khairallah, A.~A. Martin, J.~R.~I. Lee, G.~Guss, N.~P. Calta, J.~A.
  Hammons, M.~H. Nielsen, K.~Chaput, E.~Schwalbach, M.~N. Shah, M.~G. Chapman,
  T.~M. Willey, A.~M. Rubenchik, A.~T. Anderson, Y.~M. Wang, M.~J. Matthews,
  W.~E. King, Controlling interdependent meso-nanosecond dynamics and defect
  generation in metal 3d printing, Science 368~(6491) (2020) 660--665.

\bibitem{Anisimov1995}
S.~I. Anisimov, V.~A. Khokhlov, {Instabilities in laser-matter interaction},
  CRC Press, Boca Raton, 1995.

\bibitem{Lee2015}
Y.~S. Lee, W.~Zhang, {Mesoscopic simulation of heat transfer and fluid flow in
  laser powder bed additive manufacturing}, in: Solid Free Form Fabrication
  Symposium, Austin, 2015, pp. 1154--1165.

\bibitem{Leitz2018}
K.-H. Leitz, C.~Grohs, P.~Singer, B.~Tabernig, A.~Plankensteiner, H.~Kestler,
  L.~S. Sigl, {Fundamental analysis of the influence of powder characteristics
  in Selective Laser Melting of molybdenum based on a multi-physical simulation
  model}, International Journal of Refractory Metals and Hard Materials 72
  (2018) 1--8.

\bibitem{Qiu2015}
C.~Qiu, C.~Panwisawas, M.~Ward, H.~Basoalto, J.~Brooks, M.~Attallah, {On the
  role of melt flow into the surface structure and porosity development during
  selective laser melting}, Acta Materialia 96 (2015) 72--79.

\bibitem{Panwisawas2017}
C.~Panwisawas, C.~Qiu, M.~J. Anderson, Y.~Sovani, R.~P. Turner, M.~M. Attallah,
  J.~W. Brooks, H.~C. Basoalto, {Mesoscale modelling of selective laser
  melting: Thermal fluid dynamics and microstructural evolution}, Computational
  Materials Science 126 (2017) 479--490.

\bibitem{Otto2012}
A.~Otto, H.~Koch, R.~G. Vazquez, {Multiphysical Simulation of Laser Material
  Processing}, Physics Procedia 39 (2012) 843--852.

\bibitem{Yan2018}
W.~Yan, Y.~Qian, W.~Ge, S.~Lin, W.~K. Liu, F.~Lin, G.~J. Wagner, {Meso-scale
  modeling of multiple-layer fabrication process in selective electron beam
  melting: inter-layer/track voids formation}, Materials {\&} Design 141 (2018)
  210--219.

\bibitem{egorov2020}
S.~Egorov, R.~Khmyrov, A.~Korotkov, A.~Gusarov, Experimental study and modeling
  of melt pool in laser powder-bed fusion of thin walls, Procedia CIRP 94
  (2020) 372--377.

\bibitem{Gurtler2013}
F.-J. G{\"{u}}rtler, M.~Karg, K.-H. Leitz, M.~Schmidt, {Simulation of Laser
  Beam Melting of Steel Powders using the Three-Dimensional Volume of Fluid
  Method}, Physics Procedia 41 (2013) 881--886.

\bibitem{Yu2016}
G.~Yu, D.~Gu, D.~Dai, M.~Xia, C.~Ma, Q.~Shi, {On the role of processing
  parameters in thermal behavior, surface morphology and accuracy during laser
  3D printing of aluminum alloy}, Journal of Physics D: Applied Physics 49~(13)
  (2016) 135501.

\bibitem{Yuan2015}
P.~Yuan, D.~Gu, {Molten pool behaviour and its physical mechanism during
  selective laser melting of TiC/AlSi10Mg nanocomposites: simulation and
  experiments}, Journal of Physics D: Applied Physics 48~(3) (2015) 35303.

\bibitem{Tan2013}
W.~Tan, N.~S. Bailey, Y.~C. Shin, {Investigation of keyhole plume and molten
  pool based on a three-dimensional dynamic model with sharp interface
  formulation}, Journal of Physics D: Applied Physics 46~(5) (2013).

\bibitem{Tan2014}
W.~Tan, Y.~C. Shin, {Analysis of multi-phase interaction and its effects on
  keyhole dynamics with a multi-physics numerical model}, Journal of Physics D:
  Applied Physics 47~(34) (2014).

\bibitem{Kouraytem2019}
N.~Kouraytem, X.~Li, R.~Cunningham, C.~Zhao, N.~Parab, T.~Sun, A.~D. Rollett,
  A.~D. Spear, W.~Tan, {Effect of Laser-Matter Interaction on Molten Pool Flow
  and Keyhole Dynamics}, Physical Review Applied 11~(6) (2019) 1--16.

\bibitem{Korner2011}
C.~K{\"{o}}rner, E.~Attar, P.~Heinl, {Mesoscopic simulation of selective beam
  melting processes}, Journal of Materials Processing Technology 211~(6) (2011)
  978--987.

\bibitem{Korner2013}
C.~K{\"{o}}rner, A.~Bauerei{\ss}, E.~Attar, {Fundamental consolidation
  mechanisms during selective beam melting of powders}, Modelling and
  Simulation in Materials Science and Engineering 21~(8) (2013) 85011.

\bibitem{Markl2015}
M.~Markl, R.~Ammer, U.~R{\"{u}}de, C.~K{\"{o}}rner, {Numerical investigations
  on hatching process strategies for powder-bed-based additive manufacturing
  using an electron beam}, International Journal of Advanced Manufacturing
  Technology 78~(1-4) (2015) 239--247.

\bibitem{Morris1997}
J.~P. Morris, P.~J. Fox, Y.~Zhu, {Modeling low Reynolds number incompressible
  flows using SPH}, Journal of Computational Physics 136~(1) (1997) 214--226.

\bibitem{proell2020phase}
S.~D. Proell, W.~A. Wall, C.~Meier, {On phase change and latent heat models in
  metal additive manufacturing process simulation}, Advanced Modeling and
  Simulation in Engineering Sciences 7 (2020) 1--32.

\bibitem{Liu2010}
M.~B. Liu, G.~R. Liu, {Smoothed particle hydrodynamics (SPH): An overview and
  recent developments}, Archives of Computational Methods in Engineering 17~(1)
  (2010) 25--76.

\bibitem{Monaghan2005}
J.~J. Monaghan, Smoothed particle hydrodynamics, Reports on progress in physics
  68~(8) (2005) 1703--1759.

\bibitem{Quinlan2006}
N.~J. Quinlan, M.~Basa, M.~Lastiwka, {Truncation error in mesh-free particle
  methods}, International Journal for Numerical Methods in Engineering 66~(13)
  (2006) 2064--2085.

\bibitem{Adami2012}
S.~Adami, X.~Y. Hu, N.~A. Adams, {A generalized wall boundary condition for
  smoothed particle hydrodynamics}, Journal of Computational Physics 231~(21)
  (2012) 7057--7075.

\bibitem{Adami2013}
S.~Adami, X.~Y. Hu, N.~A. Adams, {A transport-velocity formulation for smoothed
  particle hydrodynamics}, Journal of Computational Physics 241 (2013)
  292--307.

\bibitem{Fuchs2020}
S.~L. Fuchs, C.~Meier, W.~A. Wall, C.~J. Cyron, {A novel smoothed particle
  hydrodynamics and finite element coupling scheme for fluid-structure
  interaction: the sliding boundary particle approach}, submitted for
  publication, arXiv:2010.09526 (2020).

\bibitem{Monaghan1983}
J.~J. Monaghan, R.~A. Gingold, {Shock simulation by the particle method SPH},
  Journal of Computational Physics 52~(2) (1983) 374--389.

\bibitem{Berger2005}
M.~Berger, M.~Aftosmis, S.~Muman, {Analysis of slope limiters on irregular
  grids}, in: 43rd AIAA Aerospace Sciences Meeting and Exhibit, 2005, p. 490.

\bibitem{Monaghan2006}
J.~J. Monaghan, {Smoothed particle hydrodynamic simulations of shear flow},
  Monthly Notices of the Royal Astronomical Society 365~(1) (2006) 199--213.

\bibitem{Gounley2016}
J.~Gounley, G.~Boedec, M.~Jaeger, M.~Leonetti, {Influence of surface viscosity
  on droplets in shear flow}, Journal of Fluid Mechanics 791 (2016) 464.

\bibitem{Cleary1999}
P.~W. Cleary, J.~J. Monaghan, {Conduction Modelling Using Smoothed Particle
  Hydrodynamics}, Journal of Computational Physics 148~(1) (1999) 227--264.

\bibitem{Cleary1998}
P.~W. Cleary, {Modelling confined multi-material heat and mass flows using
  SPH}, Applied Mathematical Modelling 22~(12) (1998) 981--993.

\bibitem{Chen2001}
J.~K. Chen, J.~E. Beraun, C.~J. Jih, {A corrective smoothed particle method for
  transient elastoplastic dynamics}, Computational Mechanics 27~(3) (2001)
  177--187.

\bibitem{Bonet1999}
J.~Bonet, T.~S. Lok, {Variational and momentum preservation aspects of Smooth
  Particle Hydrodynamic formulations}, Computer Methods in Applied Mechanics
  and Engineering 180~(1-2) (1999) 97--115.

\bibitem{Ma2011}
C.~Ma, D.~Bothe, {Direct numerical simulation of thermocapillary flow based on
  the Volume of Fluid method}, International Journal of Multiphase Flow 37~(9)
  (2011) 1045--1058.

\bibitem{rayleigh1879capillary}
L.~Rayleigh, et~al., On the capillary phenomena of jets, Proc. R. Soc. London
  29~(196-199) (1879) 71--97.

\bibitem{gusarov2010}
A.~Gusarov, I.~Smurov, Modeling the interaction of laser radiation with powder
  bed at selective laser melting, Physics Procedia 5 (2010) 381--394.

\bibitem{King2014}
W.~E. King, H.~D. Barth, V.~M. Castillo, G.~F. Gallegos, J.~W. Gibbs, D.~E.
  Hahn, C.~Kamath, A.~M. Rubenchik, Observation of keyhole-mode laser melting
  in laser powder-bed fusion additive manufacturing, Journal of Materials
  Processing Technology 214~(12) (2014) 2915 -- 2925.

\bibitem{cunningham2019keyhole}
R.~Cunningham, C.~Zhao, N.~Parab, C.~Kantzos, J.~Pauza, K.~Fezzaa, T.~Sun,
  A.~D. Rollett, Keyhole threshold and morphology in laser melting revealed by
  ultrahigh-speed x-ray imaging, Science 363~(6429) (2019) 849--852.

\end{thebibliography}
\end{document}